\documentclass[prc,aps,twocolumn,showpacs,superscriptaddress]{revtex4-1}

\usepackage{array}
\usepackage{float}
\usepackage{amsmath}
\usepackage{graphicx}
\usepackage{epsfig}
\usepackage{amssymb}
\usepackage{color}
\usepackage{subfigure}
\usepackage{sidecap}
\usepackage{txfonts}
\usepackage{sidecap}
\usepackage{hyperref}
\usepackage{calc}
\usepackage[normalem]{ulem}
\usepackage[colorinlistoftodos]{todonotes}

\usepackage{cancel}

\providecommand{\aap}{Astron.\ Astrophys.}
\providecommand{\apjl}{Astrophys.\ J.}

\providecommand{\physrep}{Phys.\ Rep.}
\providecommand{\pasa}{Publ.\ Astron.\ Soc.\ Austr.}
\providecommand{\mnras}{Mon.\ Not.\ Roy.\ Astron.\ Soc.}
\providecommand{\nphysa}{Nucl.\ Phys.\ A}

\begin{document}

\title{Medium modifications for light and heavy nuclear clusters in simulations of core collapse supernovae -- Impact on equation of state and weak interactions}

\author{Tobias Fischer}\email{tobias.fischer@uwr.edu.pl}
\affiliation{Institute for Theoretical Physics, University of Wroc{\l}aw, plac Maksa Borna 9, 50-204 Wroc{\l}aw, Poland}

\author{Stefan Typel}
\affiliation{Technische Universit{\"a}t Darmstadt, Fachbereich Physik, Institut f\"{u}r Kernphysik, Schlossgartenstra{\ss}e 9, 64289 Darmstadt, Germany }
\affiliation{GSI Helmholtzzentrum f\"ur Schwerionenforschung, Theorie, Planckstra{\ss}e~1, 64291 Darmstadt, Germany}

\author{Gerd~R{\"o}pke}
\affiliation{Institut f{\"u}r Physik, Universit{\"a}t Rostock, Albert-Einstein-Stra{\ss}e 23 - 24, 18059 Rostock, Germany}

\author{Niels-Uwe~F.~Bastian}
\affiliation{Institute for Theoretical Physics, University of Wroc{\l}aw, plac Maksa Borna 9, 50-204 Wroc{\l}aw, Poland}

\author{Gabriel~Mart{\'i}nez-Pinedo}
\affiliation{GSI Helmholtzzentrum f\"ur Schwerionenforschung, Theorie, Planckstra{\ss}e~1, 64291 Darmstadt, Germany}
\affiliation{Technische Universit{\"a}t Darmstadt, Fachbereich Physik, Institut f\"{u}r Kernphysik, Schlossgartenstra{\ss}e 9, 64289 Darmstadt, Germany }
\affiliation{Helmholtz Forschungsakademie Hessen f{\"u}r FAIR, GSI Helmholtzzentrum f{\"u}r Schwerionenforschung, Planckstra{\ss}e~1,  64291 Darmstadt, Germany}

\begin{abstract}
  The present article investigates the role of heavy nuclear clusters
  and weakly bound light nuclear clusters based on a newly developed
  equation of state for core collapse supernova studies. A novel
  approach is brought forward for the description of nuclear clusters,
  taking into account the quasiparticle approach and continuum
  correlations. It demonstrates that the commonly employed nuclear
  statistical equilibrium approach, based on non-interacting
  particles, for the description of light and heavy clusters becomes
  invalid for warm nuclear matter near the saturation density. This has
  important consequences for studies of core collapse supernovae.
  To this end, we implement this nuclear equation of state provided
  for arbitrary temperature, baryon density and isospin asymmetry,
  to spherically symmetric core collapse supernova simulations in 
  order to study the impact on the dynamics as well as on the neutrino emission. For the inclusion
  of a set of weak processes involving light clusters the rate
  expressions are derived, including medium modifications at the
  mean field level. A substantial impact from the inclusion of a
  variety of weak reactions involving light clusters on the
  post bounce dynamics nor on the neutrino emission could not be
  found.
\end{abstract}

\received{\today}

\maketitle

%%%%%%%%%%%%%%%%%%%%%%%%%%%%%%%%%%%%%%%%%%%%%%%%%%%%%%%%%%%%%%%%%%%%%%%%%%%%%%%%%%%%%%%%%%
\section{Introduction}
\label{sec:intro}
A core collapse supernova (SN) is associated with the gravitational collapse of a massive star's core at the end of its life, due to the loss of the dominant pressure from the degenerate electron gas. This is caused by electron captures on protons bound in nuclei and is accompanied by the photodisintegration of heavy nuclei. A proto-neutron star (PNS) forms at the center when the density exceeds nuclear saturation density during the stellar core contraction. Then, the short-range repulsive nuclear interaction in homogeneous nuclear matter halts the collapse, a hydrodynamics shock wave forms and the core bounces back. This instant is defined when the maximum central density is reached, corresponding to the moment just before the shock breakout. The shock wave stalls during its initial propagation out of the core due to the dissociation of still infalling heavy nuclei and the deleptonization associated with $\nu_e$-losses when the shock wave propagates across the neutrinosphere. The revival of the stalled bounce shock, i.e. the SN  problem, is related to the liberation of energy from the central PNS into the low-density layer behind the shock (for a review, c.f. Ref.~\cite{Janka:2007}). The shock revival to increasingly larger radii defines the onset of the SN explosion.

The SN conditions are wide spread, see Fig.~1(a) in
Ref.~\cite{Fischer:2017}, featuring temperatures of
$T\simeq0-100$~MeV, restmass densities of
$\rho=10^0-10^{15}$~g~cm$^{-3}$ and isospin asymmetry given by the
charge density, or equivalent the proton abundance $Y_p$ ranging from
$Y_p=0$ (neutron matter) to $Y_p=1$ (proton matter). Note that the
proton fraction equals the electron fraction, $Y_p=Y_e$, in the
absence of other leptonic charges.
For recent comprehensive reviews about core collapse SN phenomenology, see 
Refs.~\cite{Janka:2012,Mueller:2016}, discussing also the role of the equation of state (EOS)
in SN simulations at the various different regimes encountered~\cite{Fischer:2017}.

The EOS at the suprasaturation density,  $\rho_0=2.5\times 10^{14}$~g~cm$^{-3}$ 
(equivalent the nucleon number density, $n_0=0.15$~fm$^{-3}$), as well as high temperatures is 
often obtained from nuclear energy density functionals. These are often derived in a mean-field 
approximation for the interacting nucleons. At subsaturation densities and temperatures below few 
tens of MeV, inhomogeneous matter can be found where nuclear clusters as well as unbound 
nucleons are present in addition to pasta
phases~\cite{Schuetrumpf.Martinez-Pinedo.ea:2019}. Even neglecting
pasta phases, the description of the nuclear medium is particularly challenging.
The commonly employed approaches for SN EOS are the single-nucleus approximation
based on the liquid-drop model (see, e.g., Refs.~\cite{Lattimer:1991nc,Schneider:2017} and 
references therein), the Thomas-Fermi approach of Ref.~\cite{Shen:1998by} and the modified 
nuclear statistical equilibrium (NSE) of Ref.~\cite{Hempel:2009mc} (see also 
Ref.~\cite{Furusawa:2011}). Only the latter allows for the detailed nuclear composition
to be taken into account explicitly. It is based on several thousand nuclear
species with tabulated (and partly calculated) nuclear masses, as well as a theoretical
framework handling excited states which must be included at finite temperatures. In 
particular, about the role of light nuclear clusters, such as deuteron $^2$H, has long
been speculated \cite{Furusawa:2013}. The potential role of additional clusters, such
as triton, $^3$H, and helium-3, $^3$He, has been pointed out in Ref.~\cite{Fischer:2017},
also its particular impact on the neutrino response \cite{Arcones:2008}. Recently, even
more {\em exotic} neutron rich isotopes have been studied in the context of core 
collapse SNe in Ref.~\cite{Yudin:2019} based on the simplistic NSE model of 
Ref.~\cite{Hempel:2009mc}, which is based on nuclei treated as non-interacting
Boltzmann particles.

Despite the success of this NSE approach, there are a number of
caveats. Whereas the NSE is well suited to describe nuclear matter at
low densities and high temperatures, the account of medium effects is
necessary when going to higher densities ($\gtrsim 10^{12}$~g~cm$^{-3}$).
One of the most important issues is the transition to  homogeneous matter
when approaching the saturation density. The need to take density effects
(medium modifications of the cluster properties) into account was demonstrated
by laboratory experiments \cite{Qin:2012,Pais:2020}. A quantum-statistical approach
(see \cite{Roepke:2020} and references given therein), has been worked
out which describes a generalized virial expansion to account for 
correlations in the continuum. Self-energy and Pauli-blocking contributions
are obtained from an in-medium few-nucleon Schr\"{o}dinger equation. In
the low-density region, the relation to the concept of excluded volume
has been demonstrated in Ref.~\cite{Hempel:2015b}, and a generalized
relativistic density-functional (gRDF) approach \cite{Typel:2009sy} which
considers the light elements as new quasiparticles coupled to the mesonic
fields. The semi-empirical approaches, i.e. excluded volume and gRDF,
proved to be successful in describing laboratory experiments in the density
region up to one fifth of the saturation density. A modification of the 
relativistic mean-field approach considering in-medium effects via an
effective cluster-meson coupling has been proposed \cite{Pais:2018,Pais:2019},
where light clusters $A \le 12$ have been taken into account.

It was demonstrated in Sec.~4 of Ref.~\cite{Fischer:2017}, when comparing
the modified NSE abundances (taking the excluded volume into account)
for the light nuclear clusters and unbound nucleons with those
obtained within the quantum statistical (QS) model of
Ref.~\cite{Roepke:2009} and Ref.~\cite{Roepke:2011}, that the
NSE model of Ref.~\cite{Hempel:2009mc} significantly overestimates the abundances 
of all light clusters in the density region $\rho\ge 10^{12}$~g~cm$^{-3}$. Also the
properties of the unbound nucleons are affected substantially. In
order to quantify this impact in simulations of core collapse SNe, the
present work expands the previous analysis, by means of developing a
novel fully temperature and isospin asymmetry dependent SN EOS based
on the gRDF model of Ref.~\cite{Pais:2016fdp}, including self-consistently
the description of light nuclear clusters (defined as nuclei with
charge $Z\leq 2$) and heavy clusters ($Z>2$) as well as the transition
to homogeneous matter. The gRDF approach implements medium modified
nuclear binding energies (see Ref.~\cite{Roepke:2020}) that are derived
to match the QS approach, as well as continuum correlations described
by the virial expansion, which ensures the appropriate transition to
homogeneous matter. This has important consequences for the SN
dynamics and neutrino emission. To study systematically the
differences obtained in comparison to the NSE EOS of
Ref.~\cite{Hempel:2009mc}, henceforth denoted as HS, we implement the gRDF
EOS into the SN model {\tt AGILE-BOLTZTRAN} and simulate the
post bounce evolution in spherical symmetry. We implement the same
density dependent mean-field parametrisation (denoted as DD2) of
Ref.~\cite{Typel:2009sy}, for both EOS henceforth denoted as gRDF(DD2) and
HS(DD2) (The HS EOS catalogue is provided by
CompOSE~\cite{compose} and has been subject to extensive
comparison studies, c.f., Ref.~\cite{Oertel:2017} and references therein).

Neutrinos emitted from a core collapse SN contain a broad
spectrum of information, e.g., they probe the EOS deep inside the
SN core which is otherwise hidden. The future neutrino
detection from the next galactic event will shed light not only on
details of the yet controversially discussed explosion mechanism, it
may also reveal properties of matter at conditions which are currently
inaccessible in nuclear physics experiments. It is of paramount
interest to predict reliable neutrino luminosities and spectra, as
well as their evolution, for such events. Of particular importance is
the role of the hot and dense nuclear medium. Besides weak processes
involving the unbound nucleons, the role of heavy nuclear clusters has
long been studied, i.e. for nuclear electron capture rates~\cite{Langanke:2003b}
as well as neutrino-nucleus scattering (see Ref.~\cite{Langanke:2007ua} and references therein)
and nuclear (de)excitations \cite{Fuller:1991,Fischer:2013}.
The previous implementation of weak charged current reactions with light
nuclear clusters in core collapse SN simulations has so far still been
approximate \cite{Furusawa:2013}. The gRDF(DD2) EOS, subject to
the present paper, contains the detailed composition, including all
light clusters. This allows us to extend the previous attempts and include
a variety of weak processes, involving the hydrogen isotopes with atomic mass
numbers $A=2-7$ and the helium isotopes with $A=3-7$, into the
simulations of the SN post bounce evolution. In particular, we discuss
the significance of exotic, neutron-rich H (e.g., $^4$H) and He (e.g.,
$^5$He) isotopes near the saturation density as proposed recently by
Ref.~\cite{Yudin:2019} within the simple NSE model of Ref.~\cite{Hempel:2009mc}.
This approach overestimates the abundances of these light clusters as shown recently
in Ref.~\cite{Roepke:2020} within the QS formalism. The account of the
light element isotopes leads to a reduction of the neutrino
luminosities and an enhancement of the spectral differences between
$\nu_e$ and $\bar\nu_e$. However, a substantial impact from the
inclusion of a variety of weak reactions involving light clusters could not be
found, neither on the post bounce dynamics nor on the neutrino emission.

The manuscript is organized as follows. In Sec.~\ref{sec:SN-model} the SN model 
is briefly reviewed. The newly developed gRDF(DD2) EOS is introduced in Sec.~\ref{sec:eos}, 
in comparison to the modified NSE EOS, while in Sec.~\ref{sec:SN-sim} simulations of 
core collapse SN are discussed, comparing the reference simulation with gRDF(DD2) 
EOS and  the NSE approach. The analysis is extended in Sec.~\ref{sec:v-reactions} 
introducing weak reactions with light clusters and performing additional SN simulations. 
The manuscript closes with the summary in Sec.~\ref{sec:summary}.
\begin{table}
\centering
\caption{Standard set of weak reactions considered in this work, including references.}
\begin{tabular}{ccc}
\hline
\hline
& Weak process & Reference \\
\hline
1 & $e^- + p \rightleftarrows n + \nu_e$ & \cite{Fischer:2020} \\ 
2 & $e^+ + n \rightleftarrows p + \bar\nu_e$ & \cite{Fischer:2020} \\
3 & $n \rightleftarrows p + e^- + \bar\nu_e$ & \cite{Fischer:2020} \\
4 & $e^- + (A,Z) \rightleftarrows (A,Z-1) + \nu_e$ & \cite{Juodagalvis:2010} \\
5 & $\nu + N \rightleftarrows \nu' + N$ & \cite{Mezzacappa:1993gm,Horowitz:2001xf} \\
6 & $\nu + (A,Z) \rightleftarrows \nu' + (A,Z)$ & \cite{Mezzacappa:1993gm} \\
7 & $\nu + e^\pm \rightleftarrows \nu' + e^\pm$ & \cite{Mezzacappa:1993gx} \\
8 & $e^- + e^+ \rightleftarrows  \nu + \bar{\nu}$ & \cite{Bruenn:1985en} \\
9 & $N + N \rightleftarrows  \nu + \bar{\nu} + N + N $ & \cite{Hannestad:1997gc,Fischer:2016a} \\
10 & $\nu_e + \bar\nu_e \rightleftarrows  \nu_{\mu/\tau} + \bar\nu_{\mu/\tau}$ & \cite{Buras:2005rp,Fischer:2009} \\
11 & $\nu + \bar\nu + (A,Z) \rightleftarrows (A,Z)^*$ & \cite{Fuller:1991,Fischer:2013} \\
\hline
\end{tabular}
\\
$\nu=\{\nu_e,\bar{\nu}_e,\nu_{\mu/\tau},\bar{\nu}_{\mu/\tau}\}$ and $N=\{n,p\}$
\label{tab:nu-reactions}
\end{table}
%

%%%%%%%%%%%%%%%%%%%%%%%%%%%%%%%%%%%%%%%%%%%%%%%%%%%%%%%%%%%%%%%%%%%%%%%%%%%%%%%%%%%%%%%%%%
\section{Supernova model}\label{sec:SN-model}
The SN model employed in this study, {\tt AGILE-BOLTZTRAN}, is based on general relativistic 
neutrino-radiation hydrodynamics in spherical symmetry with three-flavor Boltzmann neutrino 
transport \cite{Mezzacappa:1993gm,Mezzacappa:1993gn,Liebendoerfer:2004}. It features a 
Lagrangian mass mesh with an adaptive mesh refinement method, originally developed in 
Ref.~\cite{Liebendoerfer:2002} and updated in Ref.~\cite{Fischer:2009}. The complete set of 
standard  weak reactions considered in this work can be found in Table~\ref{tab:nu-reactions}, 
including the references. Here we use a neutrino discretization in terms of 36 energy bins of 
$E_\nu\in [0.5,300]$~MeV and 6 momentum scattering angles, $\mu=\cos\theta\in\{-1,+1\}$, for the 
neutrino flavors $\{\nu_e,\bar\nu_e,\nu_{\mu/\tau},\bar\nu_{\mu/\tau}\}$. The implicit method solving 
the Boltzmann equation on an adaptive Lagrangian mass mesh has been compared with other 
commonly employed neutrino-transport schemes, e.g., the multi-group flux limited diffusion 
approximation \cite{Liebendoerfer:2004}, the variable Eddington factor technique 
\cite{Liebendoerfer:2005a} and with M1 schemes \cite{OConnor:2018}, resulting in good qualitative 
agreement. {\tt AGILE-BOLTZTRAN} has a flexible EOS-module that can handle many currently 
available baryon EOS 
\cite{Lattimer:1991nc,Shen:1998gg,Hempel:2009mc,Hempel:2012,Steiner:2013}.
For conditions that correspond to the non-NSE regime at low temperatures, precisely below 
$T=0.45$~MeV, we switch to a baryon EOS based on the ideal gas of silicon and sulphur nuclei. It 
is meant to resemble the remaining silicon--sulphur layer of the progenitor star. In addition to the 
baryons, contributions from $e^\pm$, photons and Coulomb are added \cite{Timmes:1999}.

Particular focus has been devoted to the consistent description of the charged current processes, reactions (1)--(3) in Table~\ref{tab:nu-reactions}, and the nuclear EOS (for details, see Ref.~\cite{Reddy:1998}), with the implementation of the nucleon mean-field potentials, $U_n$ and $U_p$~\cite{MartinezPinedo:2012,Roberts:2012}. The latter are related to the scalar $\Sigma^S_N$, Eq.~\eqref{eq:app_sigmaS}, and vector parts $\Sigma^V_N$, Eq.~\eqref{eq:app_sigmaV}, of the nucleon self energies (see Appendix~\ref{sec:Appendix_gRDF} for further details), $U_N=\Sigma^V_N-\Sigma^S_N$ (see also Ref.~\cite{Hempel:2015a} and references therein). In particular, their difference, $U_n-U_p$, is determined by the nuclear symmetry energy which has a strong density dependence \cite{Fischer:2014}. 

Of special importance for the SN dynamics, including neutrino heating and cooling, are the weak processes involving the unbound nucleons. The largest inverse mean-free paths are found for the neutral current neutrino nucleon scattering processes, reactions (5) in Table~\ref{tab:nu-reactions}, and the largest energy transfer contributions originate from the charged current absorption and emission reactions (1)--(3) in Table~\ref{tab:nu-reactions}. Therefore, not only the consistent description of the weak reaction rates and EOS, including their implementation into the transport module, are of importance but also the properties of the unbound nucleons, which in turn are determined by the nuclear EOS as well. In particular, the nucleon self-energies, which are related to the nucleon effective masses and the mean-field potentials, have a direct impact on the weak rates for both, neutral- and charged current processes (for details, see also Ref.~\cite{Fischer:2020} and references therein). More precisely, at conditions of low degeneracy, all these reaction rates are proportional to the number density of the target nucleons. Hence it is important to implement the correct description of the nucleon properties which are strongly affected by the presence of nuclear clusters in the inhomogeneous nuclear matter phase. This will be further discussed and illustrated in the following section. 

\begin{figure*}
\centering
\includegraphics[width=1.75\columnwidth]{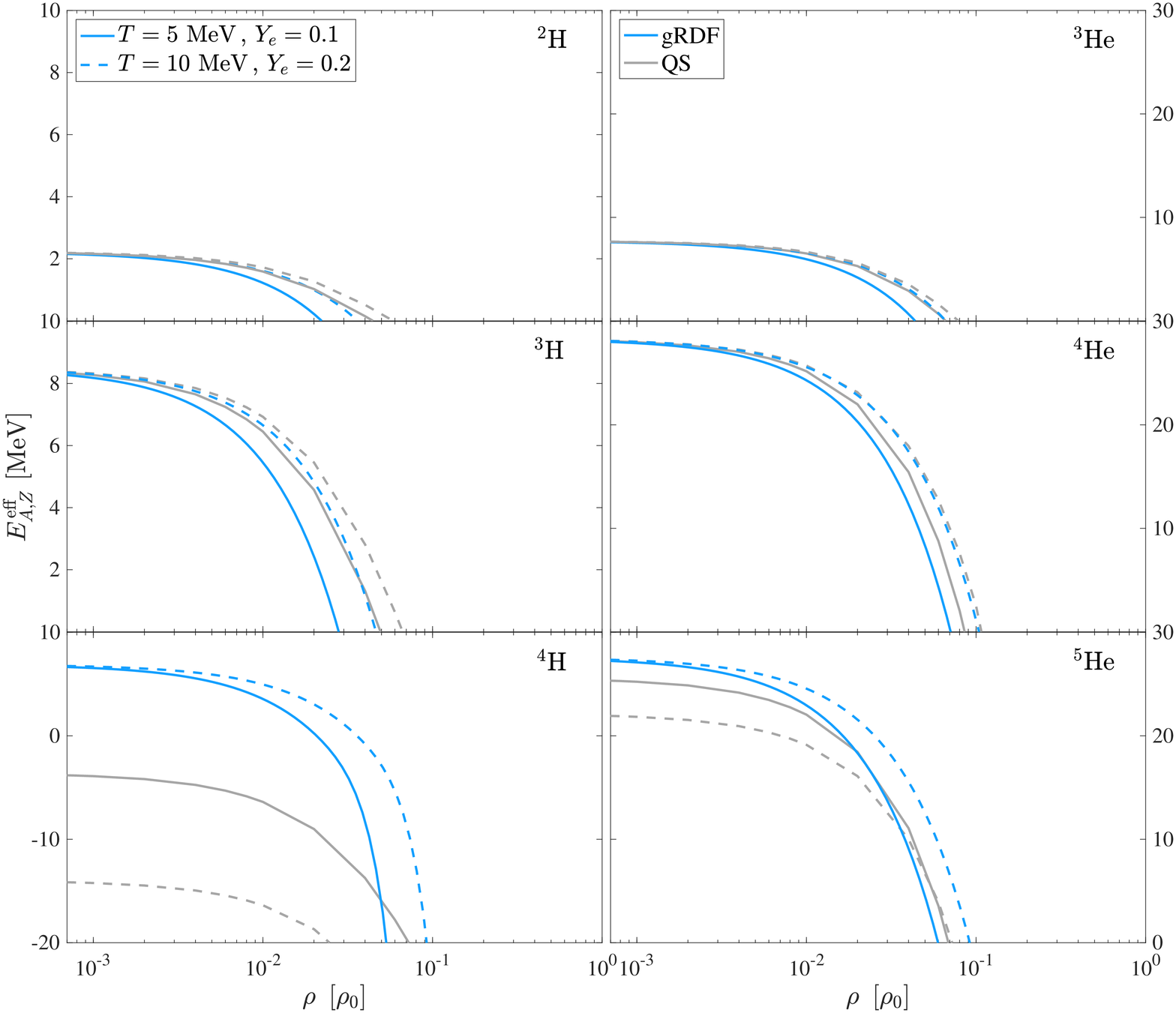}
\caption{Effective nuclear binding energy for hydrogen and helium isotopes for two representative conditions, with $T=5$~MeV and $Y_e=0.1$ (solid lines) as well as $T=10$~MeV and $Y_e=0.2$ (dashed lines), comparing the gRDF approach including a shift due to Pauli blocking (blue lines) and the QS approach \cite{Roepke:2020} (grey lines).}
\label{fig:Pauli}
\end{figure*}
\begin{figure*}
\centering
\subfigure[~$T=5$~MeV, $Y_e=0.1$]{
\includegraphics[width=0.96\columnwidth]{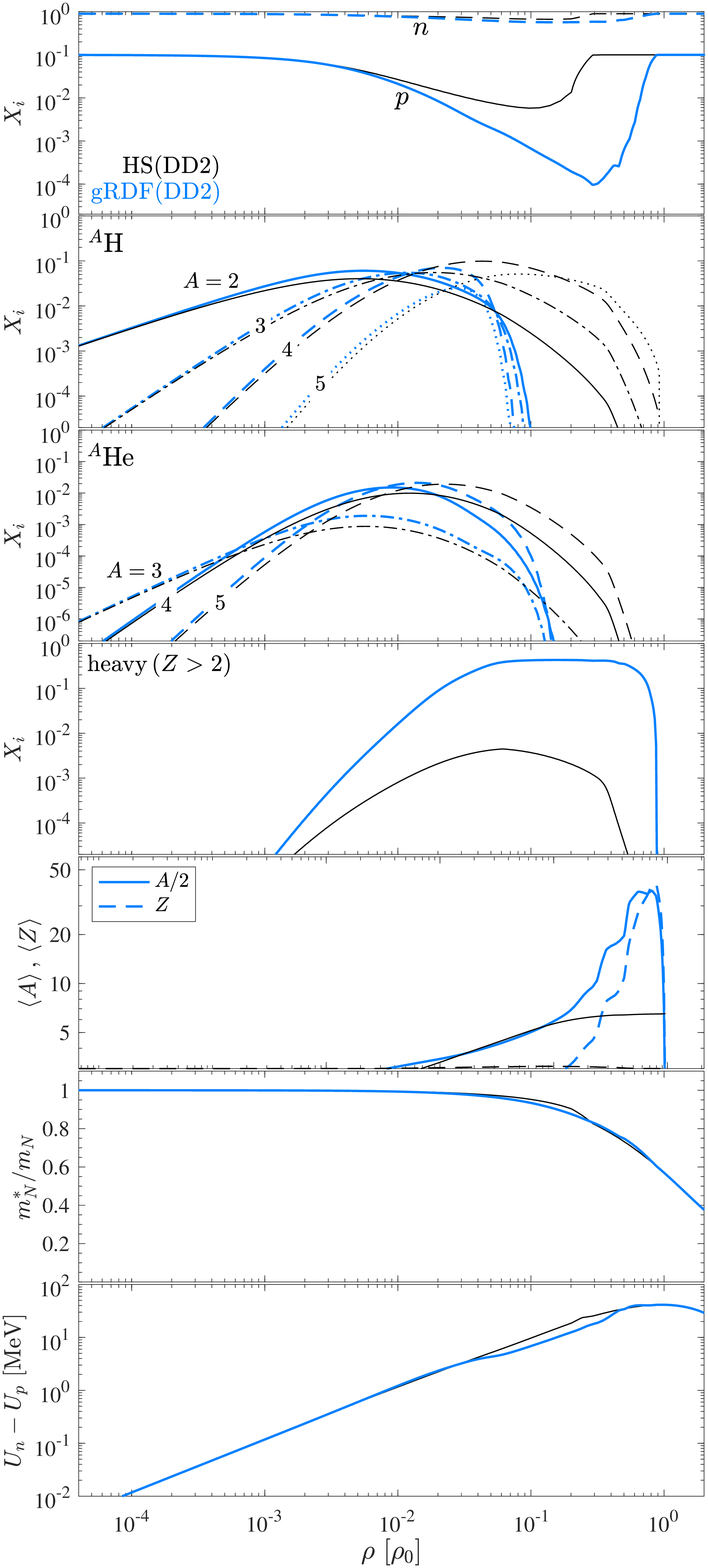}\label{fig:eos_a}}
\hfill
\subfigure[~$T=10$~MeV, $Y_e=0.2$]{
\includegraphics[width=0.96\columnwidth]{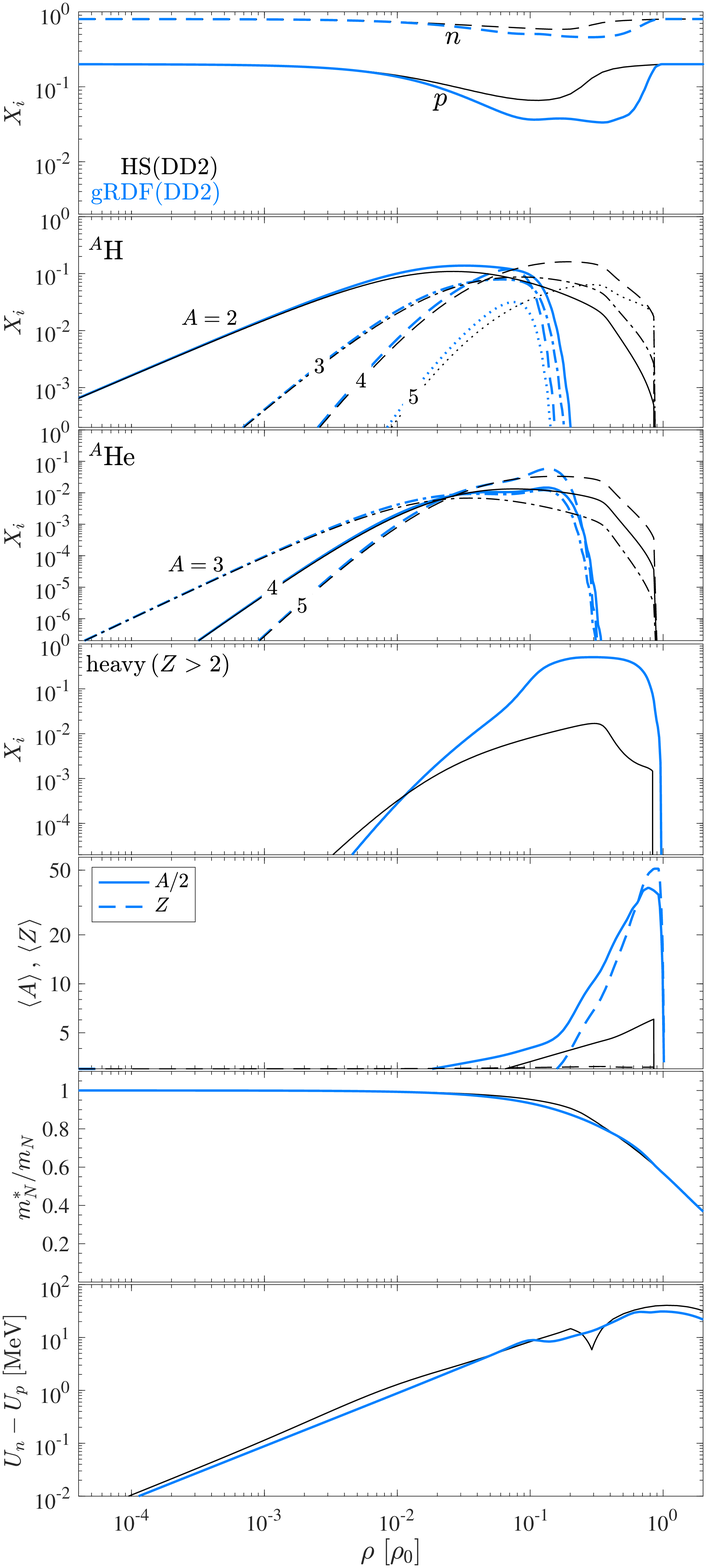}\label{fig:eos_b}}
\caption{Nuclear composition (mass fractions for neutrons and protons, hydrogen and helium isotopes, and for heavy nuclei averaging over all species with $Z>3$ as well as their average atomic mass $A$ and charge $Z$) as well as nucleon effective mass $m_N^*$ and mean-field potential difference $U_n-U_p$ as a function of the restmass density $\rho$ in units of the saturation density $\rho_0$, for a selected conditions, comparing the modified NSE model (black lines) of Ref.~\cite{Hempel:2009mc}, denoted as HS(DD2), and the gRDF(DD2) model (blue lines).}
\label{fig:eos}
\end{figure*}
%

%%%%%%%%%%%%%%%%%%%%%%%%%%%%%%%%%%%%%%%%%%%%%%%%%%%%%%%%%%%%%%%%%%%%%%%%%%%%%%%%%%%%%%%%%%
\section{Equation of state with light and heavy clusters}\label{sec:eos}
A theoretical model for the EOS used in simulations of core collapse
SNe has to provide both the thermodynamic properties and the
chemical composition of matter in a wide range of densities,
temperatures, and isospin asymmetries. In the present application, the
gRDF approach is employed which includes nucleons, nuclei, electrons,
muons, and photons as degrees of freedom. The energy density
functional was derived in the mean-field approximation from a
relativistic Lagrangian density. It describes the effective in-medium
interaction between nucleons (free and bound in clusters) by an
exchange of $\omega$, $\sigma$ and $\rho$ mesons with density
dependent couplings using the DD2 parametrisation
\cite{Typel:2009sy}. The latter was fitted to properties of finite
nuclei. The set of nuclei in the density functional comprises light
($Z \leq 2$) and heavy nuclei ($Z>2$) with experimental binding
energies of the Atomic Mass Evaluation 2016 \cite{Wang_2017}. This
table was supplemented with more exotic nuclei of unknown experimental
binding energy using the DZ31 model of Ref.~\cite{Duflo:1995ep} for their
masses. Excited states of heavy nuclei are included effectively with
temperature-dependent degeneracy factors (details can be found
in Ref.~\cite{Pais:2016fdp}). The effects of the nuclear medium on the
properties of nuclear clusters is modelled in the gRDF approach by
implementing nuclear mass shifts that lead to effective binding
energies different from their vacuum value. In addition to previous
gRDF models that considered only $^2$H, $^3$H, $^3$He, and $^4$He as
light nuclei, the present calculation incorporates additional
neutron-rich isotopes. These include the following hydrogen isotopes
$^4$H, $^5$H, $^6$H and $^7$H, as well as the following helium
isotopes $^5$He, $^6$He and $^7$He. All these exotic nuclei have
a positive nuclear binding energy in vacuum~\cite{Wang_2017}, but
some of them have a negative neutron separation energy are hence
unstable with respect to neutron emission, or $\beta$--decay as in the
case of $^6$He. Thus they can be described as resonances in the
continuum.

In order to implement these additional light clusters, including their
medium modifications, into the equation of state, the QS approach is
employed (c.f. Ref.~\cite{Roepke:2017}). The latter is based on a
many-particle theory which defines Green's functions and spectral
densities. Thermodynamic relations are derived consistently. A
few-body, in-medium Schr\"odinger equation can be given which
describes bound states as well as scattering states of the $A$-nucleon
cluster, where the influence of the nuclear medium is described by
self-energy and Pauli blocking terms. As a consequence, the bound
state energies and wave functions as well as the scattering phase
shifts of the $A$-nucleon cluster become dependent on the properties
of the medium, i.e., the temperature $T$, the density $\rho$, and the
charge density. This way, the quasiparticle concept which is
well-known for the single-nucleon states is generalized to nuclear
cluster states like the $\alpha$ particle. In
appendix~\ref{sec:Appendix_BU} the formulae are given for the virial
expansion and generalized Beth-Uhlenbeck equation, for the example of
the deuteron system with a single bound state and a continuum of
scattering states. Similarly, $^4$H and $^5$He systems can be treated
this way. In these cases, no bound states exist, and only the integral
over the scattering states remains in Eq.~\eqref{gBUd} in
Appendix~\ref{sec:Appendix_BU}. The chemical potentials contain the
the information about the density of neutrons and protons.
An effective cluster energy $E^{\rm eff}_{A,Z}$, c.f., Eq.~\eqref{eq:n2H}, can be introduced which 
depends strongly on the temperature, but also on the density. Details are discussed in 
Ref.~\cite{Roepke:2020} where analytical expressions for the $T$ and $\rho$ dependence of 
$E^{\rm eff}_{A,Z}$ for $^4$H and $^5$He are given.

The dissolution of nuclei with increasing temperature and density is
described in the gRDF model with the help of medium-dependent mass
shifts that include a small contribution from the screened Coulomb
interaction and the
strong-interaction shifts originating from Pauli blocking, as
introduced above. The concept of mass shifts replaces the
excluded-volume approach that was used in the modified NSE EOS of
Ref.~\cite{Hempel:2009mc}.
At present, while the QS approach can be applied to only a few selected light nuclear species, gRDF allows for the universal description of in-medium properties. Individual calculations of the energy shifts and continuum correlations for each isotope $\{A,Z\}$, in particular for heavy clusters with large mass numbers $A$ are presently not available. Therefore an approximate treatment applicable for arbitrary nuclei is necessary, as given by the gRDF approach. The functional form of the Pauli shift denoted as model GRDF2 in Ref.~\cite{Typel:2018wmm} was used for heavy nuclei and the functional form of the Pauli mass shifts of the nuclei $^4$H, $^5$H, and $^5$He was assumed to be identical to one of the closest stable nucleus in the same isotopic chain. Details on the theoretical formulation of the gRDF are given in Appendix \ref{sec:Appendix_gRDF}. Further details on the parametrisation of the gRDF(DD2) can be found in 
Refs.~\cite{Pais:2016fdp,Typel:2017upu,Typel:2018wmm}. 
The effective nuclear binding energies are shown in
Fig.~\ref{fig:Pauli}, for selected clusters with $Z=1$ (left panel)
and $Z=2$ (right panel). Here we compare the gRDF EOS (blue lines)
with the QS calculations (grey lines) of Ref.~\cite{Roepke:2020}. For the
bound nuclei $^2$H, $^3$H, $^3$He, $^4$He, the effective binding
energies approach the corresponding ground-state binding energy values
at low densities. A weak temperature dependence is seen in the QS
approach because of the account of scattering states leading to the
virial form of the EOS. This effect of continuum correlations becomes
small at low temperatures when states above the continuum edge are not
excited. A strong temperature dependence, illustrated at two examples,
$T=5$~MeV and $T=10$~MeV, is seen for the unbound nuclei $^4$H and
$^5$He which appear as a broad resonance above the edge of the
$^3$H--n and $^4$He--n continuum so that the contribution to the EOS
is given only by the scattering phase shifts. As a consequence, the
temperature effects are strong, as known from the virial expansion.
However, for lower temperatures on the order of $T=1$~MeV the continuum
contributions become less dominant and the effective in-medium nuclear 
binding energy for $^4$H and $^5$He rises towards the vacuum binding 
energy value, and hence gRDF and QS agree with each other at these 
conditions. Furthermore, for the important transition towards the 
regime where the light cluster dissolve, both gRDF and QS provide 
good qualitative agreement, with the slightly earlier drop of 
$E^{\rm eff}_{A,Z}$ for all light clusters within gRDF 
(see Fig.~\ref{fig:Pauli}) resulting in the dissolving of light 
clusters at slightly lower density (see Fig.~\ref{fig:eos}), 
compared to the QS framework.

None of these aspects are captured within the NSE framework
where vacuum binding energies are employed and the scattering
state contributions are ignored. Consequently, the abundances 
of bound states are generally overestimated within the NSE 
approach so that it fails to describe  the disappearance of 
bound states near the saturation density and the transition 
to homogeneous nuclear matter (Mott effect). This transition 
is a consequence of in-medium effects (Pauli blocking) and 
is modeled by the excluded volume approach as well as the 
generalized relativistic density-functional approach.

Figure~\ref{fig:eos} shows the nuclear composition of selected species at
same conditions as in Fig.~\ref{fig:Pauli} ($T = 5$~MeV, $Y_e = 0.1$, and $T = 10$~MeV,
$Y_e$ = 0.2) with respect to the density in units of the saturation density,
comparing the modified NSE EOS (thin black lines) of Ref.~\cite{Hempel:2009mc}
henceforth denoted as HS(DD2) and the gRDF(DD2) EOS of this work (thick blue lines). 
Plotted is the density range relevant for the SN evolution; very low densities
where in-medium effects are not relevant, are omitted here.
With increasing density when nuclear clusters become
abundant, the deviations between both approaches grow substantially.
There arise four major differences:
\begin{enumerate}
\item[(1)] The HS(DD2) EOS systematically overestimates the
abundances of light clusters. In particular in the region where
gRDF(DD2) predicts already the complete dissolution of all light
clusters, HS(DD2) still finds substantial abundances. This caveat
is related to the description of nuclear clusters in general within
HS(DD2), where the geometric excluded volume approach is
employed for the dissolution of clusters. It cannot distinguish
between heavy and light clusters and does not include a temperature
dependence. On the other hand, gRDF(DD2) is based on in-medium
modified binding energies for heavy and light nuclear clusters.
Consequently, the yields of all hydrogen and helium clusters drop to
zero at lower density, as illustrated in Fig.~\ref{fig:eos}, in
agreement with their effective nuclear binding energies dropping
below zero (see Fig.~\ref{fig:Pauli}).
\\
\item[(2)] Of interest is that the abundances of exotic, neutron-rich isotopes
such as the hydrogen isotopes $^4$H, $^5$H, $^6$H, but also the helium isotopes 
$^5$He, $^6$He, $^7$He, $^8$He, are strongly reduced within the gRDF(DD2) 
approach at densities above one tenth of the saturation density, in
comparison with HS(DD2). This strong reduction of the gRDF framework is in 
contrast to Ref.~\cite{Yudin:2019} where plots similar to
Fig.~\ref{fig:eos} are given, considering only the excluded volume concept
of the HS(DD2) model. This suppression is further enhanced in the QS approach
if accounting for continuum correlation as given by scattering phase shifts.
The Beth-Uhlenbeck formula is an exact expression for the second virial 
coefficient and may serve as a benchmark for any EOS in the low-density limit
\cite{Roepke:1990,Horowitz:2006,gshen2011b,voskresenskaya2012,Roepke:2013}.
Generalizations considering higher clusters and the introduction of the
quasiparticle picture need special attention to be consistent and to avoid
double counting. Calculations have been performed for the exotic nuclei
$^4$H, $^5$He (see Ref.~\cite{Roepke:2020}).
The QS approach including continuum correlations is not available at present
for arbitrary nuclei so that we focus on the gRDF(DD2) approach which provides us 
with an appropriate description of in-medium effects for arbitrary bound nuclei.
\\
\item[(3)] The in-medium nuclear binding energies for the heavy
clusters $(Z > 2)$ of gRDF(DD2) results in substantial abundances
at higher density, than for the NSE approach HS(DD2). Also the
average nuclear charge and mass numbers, $\langle Z \rangle$ and $\langle A \rangle$
respectively, are substantially different. In particular, the heavy nuclei within
gRDF(DD2) are less neutron rich than for HS(DD2), partly related
to their shift towards higher density.
\\
\item[(4)] Note the systematically lower abundance of $^4$He at low densities
for the gRDF(DD2) EOS in comparison with HS(DD2), which
is a attributed to the Coulomb shift of the heavy clusters due to electron
screening effects. Note that the latter are not included in the HS
EOS. In fact, the Coulomb shifts are larger than the Pauli shifts at
these low densities (and low temperatures). It results in an enhancement
of the in-medium nuclear binding energy for the heavy clusters.
Consequently, there are simultaneously more heavy cluster and less
$^4$He at these conditions. This effect becomes even more pronounced
at less isospin asymmetric conditions, which will become relevant
when discussing the SN simulation results below.
\end{enumerate}
\begin{figure*}
\subfigure[~At about 0.5~ms before core bounce]
{\includegraphics[width=0.99\columnwidth]{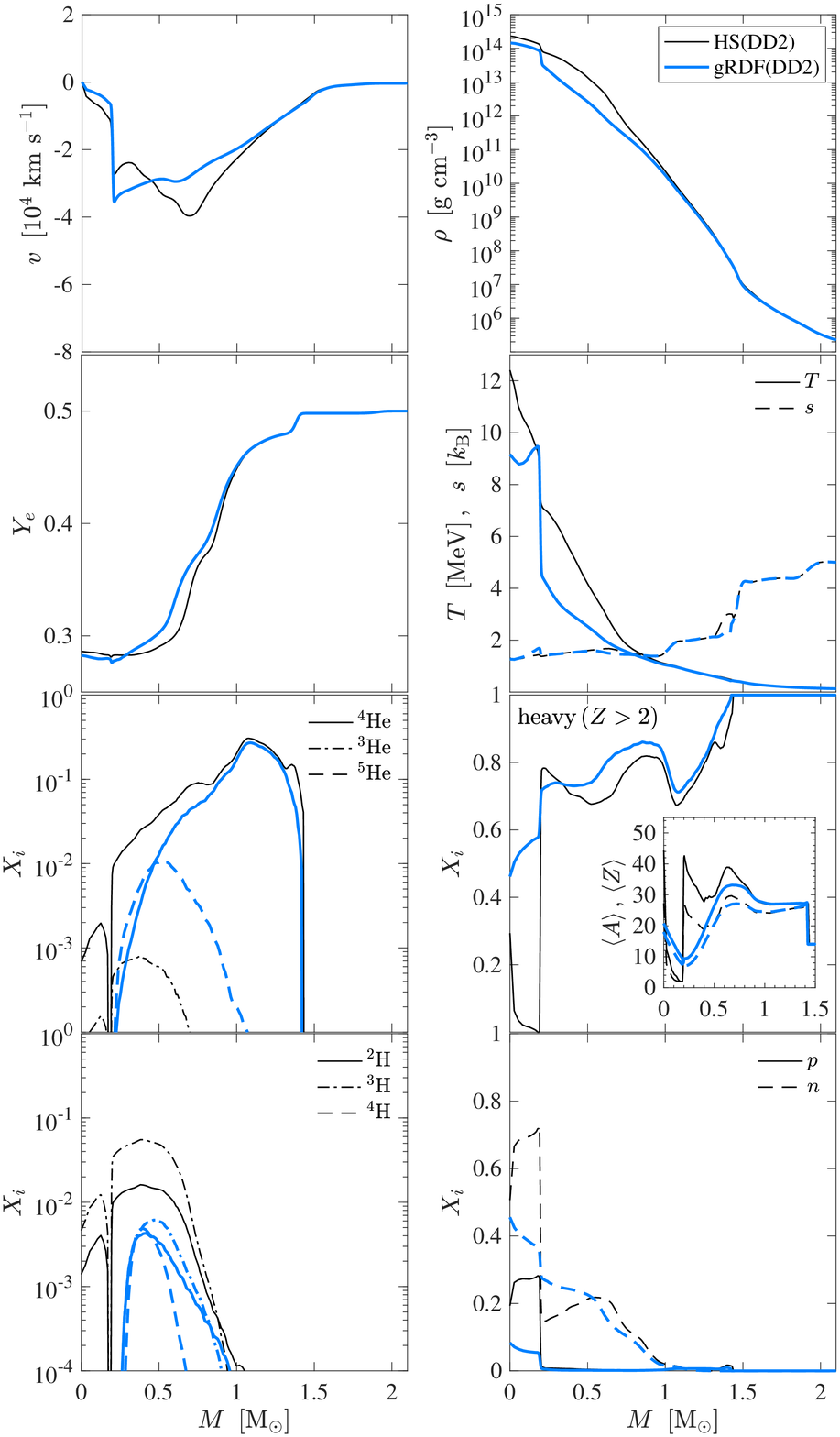}\label{fig:fullstate_collapse}}
\hfill
\subfigure[~At core bounce]
{\includegraphics[width=0.99\columnwidth]{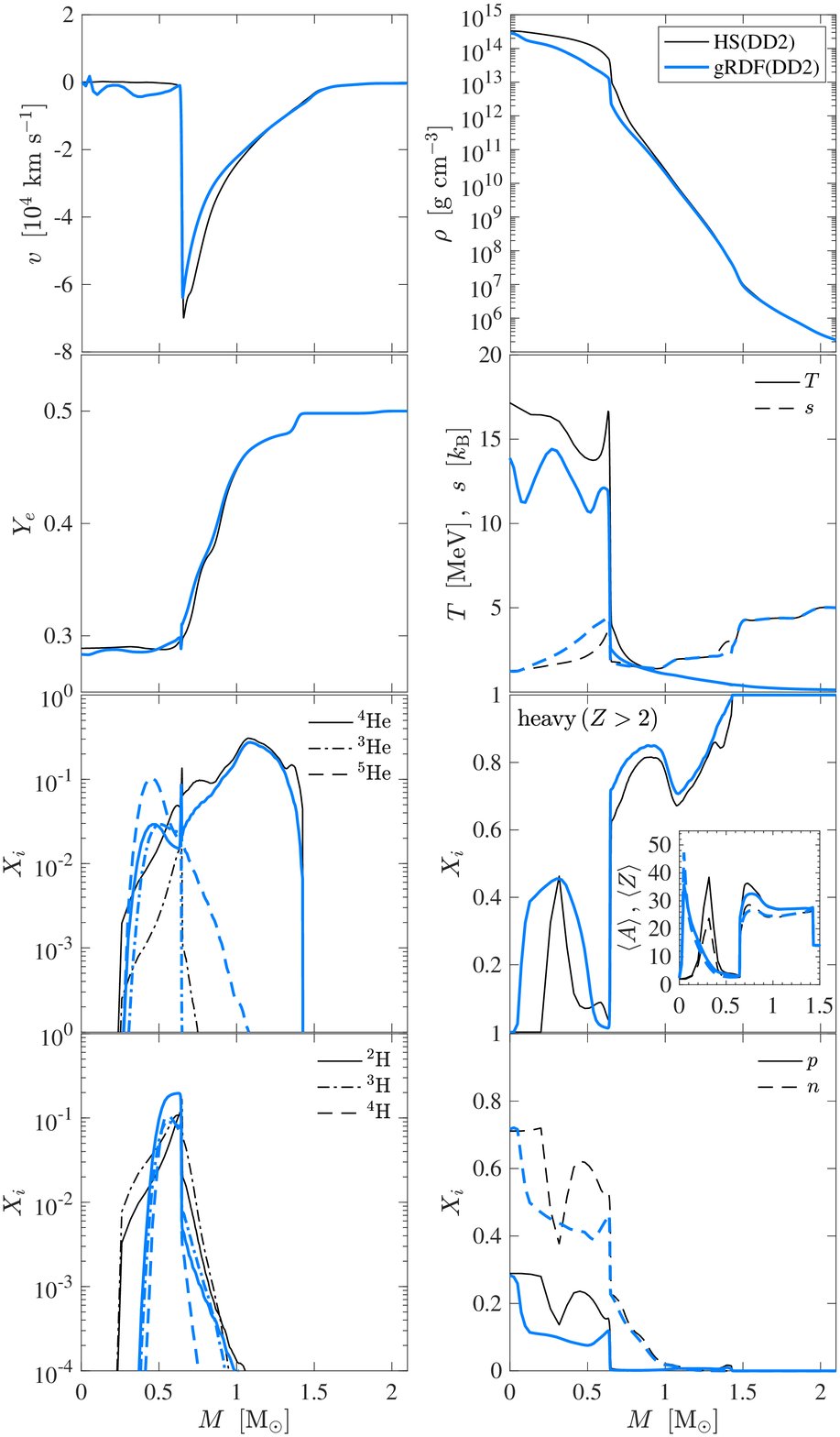}\label{fig:fullstate_bounce}}
\caption{Radial profiles of selected quantities (velocity $v$, restmass density $\rho$, electron 
fraction $Y_e$, temperature $T$ and entropy per baryon $s$, as well as various mass fraction 
$X_i$ and the heavy nuclei composition characterised by the average atomic mass $\langle A 
\rangle$ and charge $\langle Z \rangle$) as a function of the enclosed baryon mass $M$ at two 
selected times, comparing two SN simulations with the HS(DD2) EOS of 
Ref.~\cite{Hempel:2009mc} (black lines) and the newly developed gRDF(DD2) EOS (blue lines).}
\label{fig:fullstate_a}
\end{figure*}

The different description of the light and heavy nuclear clusters feeds
back to the abundances of the unbound nucleons (see the top panel in
Fig.~\ref{fig:eos}). Largest differences between HS(DD2) and gRDF(DD2) are
found in the density domain corresponding to the dissolution of the
light clusters at intermediate densities between $\rho=0.05-0.5\times\rho_0$
and the presence of heavy clusters at higher density. It results in a
substantially lower abundance of neutrons and protons at these densities.
Consequently also the nucleon self energies are modified at
the same densities (see the two bottom panels in Fig.~\ref{fig:eos}).
All this has already been studied in great detail in Ref.~\cite{Pais:2016fdp}.
It has important consequences for the SN evolution, which will be illustrated
and discussed in the next section.

We give some arguments why we think that the gRDF(DD2) EOS is more 
appropriate than the HS(DD2) EOS. The gRDF(DD2) EOS is closely related
to the QS approach and is consistent with all presently known constraints.
It matches the neutron matter calculations of chiral perturbation
theory at subsaturation density (see Refs.~\cite{Tews:2013,Krueger:2013}
and references therein) and pulsar observations. The latter refer to a neutron
star radius of about 13~km for a neutron star mass of 1.4~M$_\odot$, in
agreement with the constraint obtained from the gravitational wave analysis
of the inspiral phase of the first binary neutron star merger events
GW170817 and GW190425 \cite{Abbott:2018,Lattimer:2018,Abbott:2020},
being in agreement with the first NICER data analysis \cite{Miller:2019},
as well as a maximum neutron star mass of 2.42~M$_\odot$ in agreement
with the high-precision observations of the presently most massive pulsars
\cite{Antoniadis:2013,Cromartie:2020}.

\begin{figure*}
\subfigure[~At 10~ms after core bounce]
{\includegraphics[width=0.99\columnwidth]{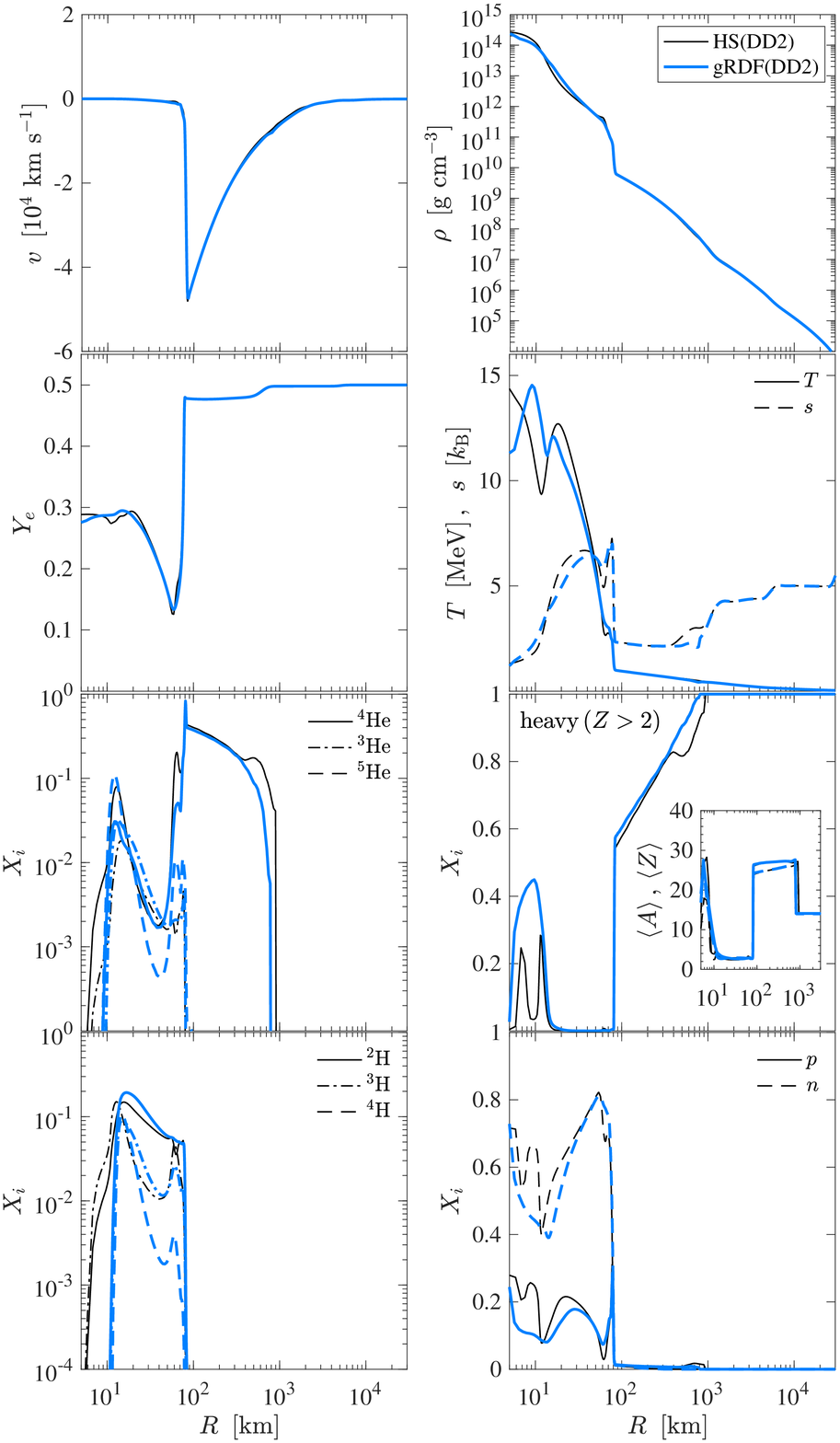}\label{fig:fullstate_001pb}}
\hfill
\subfigure[~At 100~ms after core bounce]
{\includegraphics[width=0.99\columnwidth]{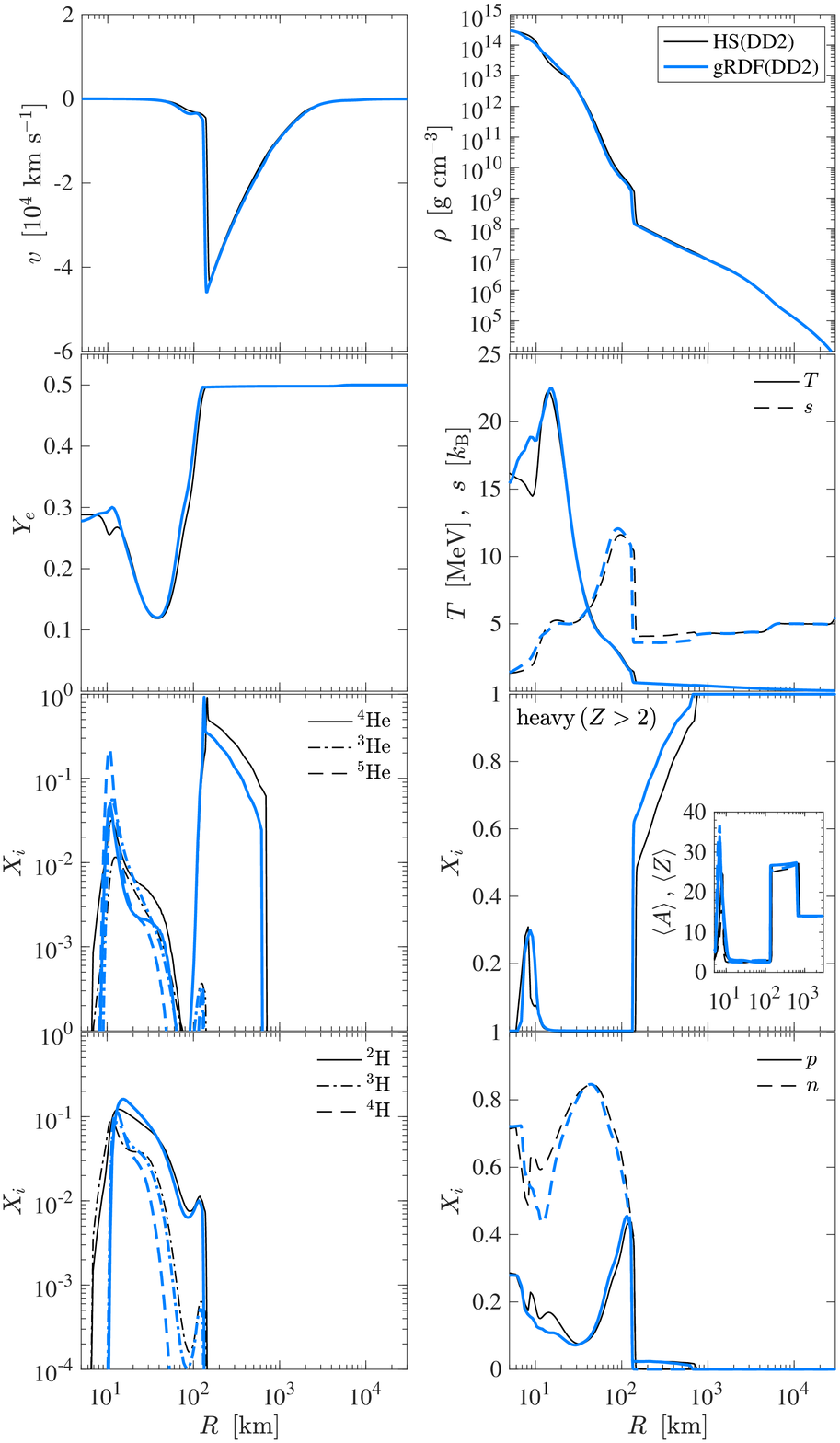}\label{fig:fullstate_010pb}}
\caption{The same quantities are shown as in Fig.~\ref{fig:fullstate_a} but with respect to the radius $R$ and for two selected post bounce times.}
\label{fig:fullstate_b}
\end{figure*}
%

%s 
\begin{figure}
\includegraphics[width=0.99\columnwidth]{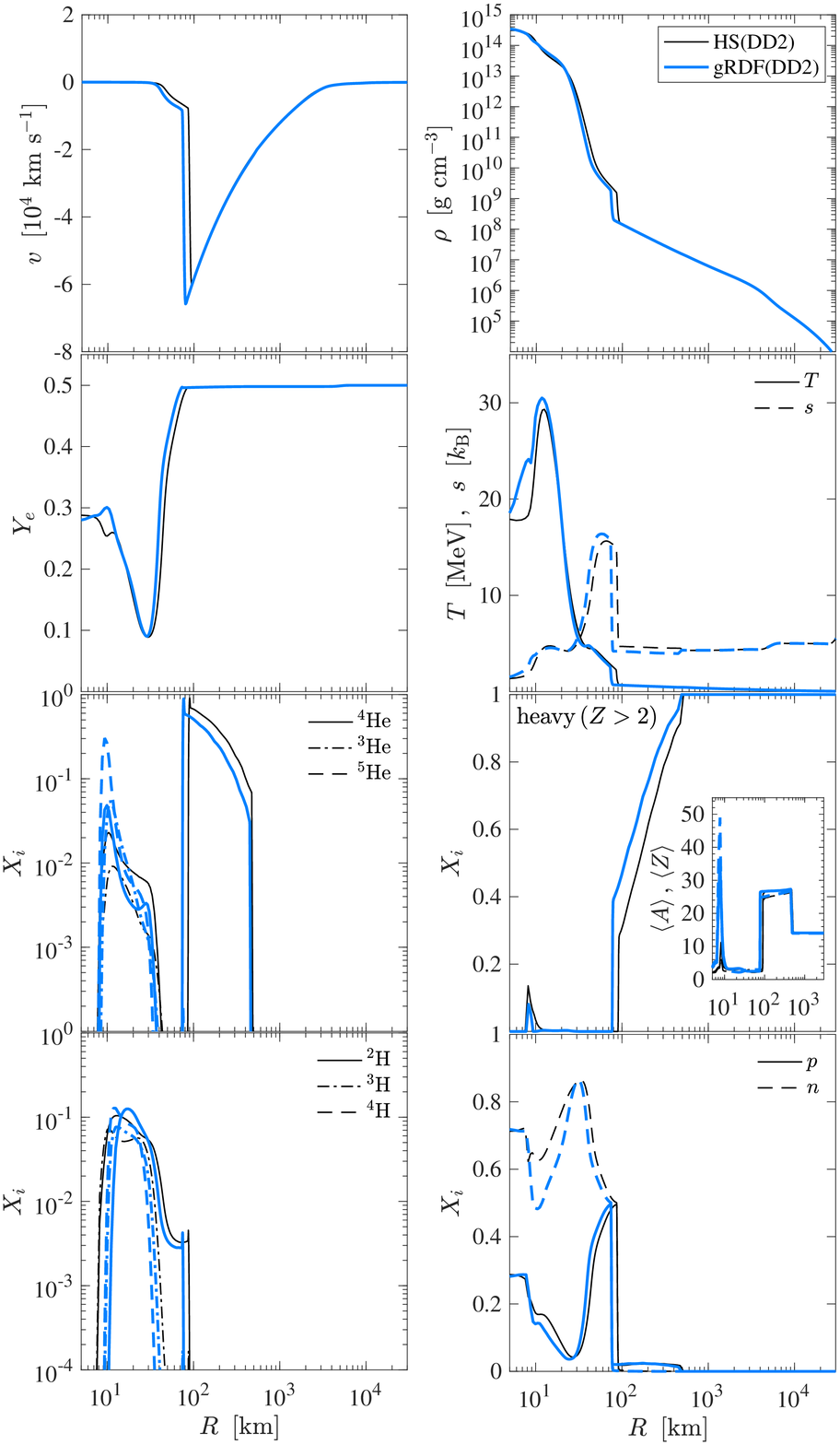}
\caption{(color online) The same quantities are shown as in Fig.~\ref{fig:fullstate_b} but at about 250~ms post bounce.}
\label{fig:fullstate_030pb}
\end{figure}
%

%%%%%%%%%%%%%%%%%%%%%%%%%%%%%%%%%%%%%%%%%%%%%%%%%%%%%%%%%%%%%%%%%%%%%%%%%%%%%%%%%%%%%%%%%%
\section{\lowercase{g}RDF EOS in core collapse SN simulations}\label{sec:SN-sim}
The core collapse SN simulations which will be discussed in the following are launched from the 18~M$_\odot$ progenitor provided by the stellar evolution series of Ref.~\cite{Woosley:2002zz}. This progenitor series has already been subject to a variety of SN EOS studies, e.g., within the failed SN branch and the subsequent black hole formation \cite{Fischer:2009}, exploring the role of the nuclear symmetry energy \cite{Fischer:2014} and the role of the stiffness/softness of the supersaturation density phase \cite{Fischer:2016b}. Here, we compare the reference simulation with the HS(DD2) EOS, which has been the foundation for our previous studies, and the newly developed and implemented gRDF(DD2) EOS. The standard set of weak processes is included (see Table~\ref{tab:nu-reactions}). SN simulations with weak processes involving light clusters will be discussed further below.

\subsection{Stellar core collapse and bounce}
The stellar core collapse evolution is governed by the presence of heavy nuclear clusters at low densities and temperatures below $T=1$~MeV. Nuclear electron captures (reaction~(4) in Table~\ref{tab:nu-reactions}) and nuclear de-excitation processes (reaction~(11) in Table~\ref{tab:nu-reactions}) give rise to neutrino losses. Since the heavy cluster component of HS(DD2) and gRDF(DD2) match at these conditions, the entire collapse phases proceeds identical. Only shortly before core bounce, when central densities close to nuclear saturation density and temperatures in excess of several MeV are reached, differences between HS(DD2) and gRDF(DD2) arise. This is illustrated in Fig.~\ref{fig:fullstate_collapse} at about 0.1~ms before core bounce. While nearly all heavy clusters are already dissolved for HS(DD2), they are still substantially abundant for the simulation with gRDF(DD2). 

The presence of a large abundance of heavy clusters at high density softens the EOS in that density domain, which has important consequences for the SN bounce dynamics. Density and temperature are notably lower for gRDF(DD2), compared to HS(DD2), in the region where heavy clusters are still abundant (see Fig.~\ref{fig:fullstate_collapse}). The central density at core bounce, shown in Fig.~\ref{fig:fullstate_bounce}, is substantially lower for gRDF(DD2), $\rho_{\rm central}=2.72\times 10^{14}$~g~cm$^{-3}$, in comparison to HS(DD2), $\rho_{\rm central}=3.32\times 10^{14}$~g~cm$^{-3}$. The energy gain due to compression work is converted into nuclear binding energy of the heavy clusters. This prevents the central density from quickly rising above the saturation density for the simulation with gRDF(DD2). A large subsaturation density domain remains where heavy clusters still exist for gRDF(DD2) (see Fig.~\ref{fig:fullstate_bounce}), featuring also lower temperatures than HS(DD2). However, the electron fraction $Y_e$ is affected only marginally. During the final core collapse phase the $Y_e$ differences can be attributed to the different nuclear composition of the heavy clusters. Already before core bounce, when the condition of weak equilibrium is obtained and the neutrino trapping regime has formed, corresponding to the high-temperature region of the collapsing stellar core where $T\gtrsim 5$~MeV, the slight differences of the central $Y_e$ obtained are due to the different density and temperature profiles, where lower density and temperature for gRDF(DD2) feature lower $Y_e$ illustrated in Fig.~\ref{fig:fullstate_bounce} at core bounce.

As expected from Ref.~\cite{Pais:2016fdp} and the analysis in
Sec.~\ref{sec:eos}, the abundances of all light clusters are
substantially overestimated for HS(DD2) in the region where they
already disappear within gRDF(DD2) towards high density. This aspect
is particularly important for the newly introduced neutron rich
hydrogen and helium isotopes, e.g., $^4$H and $^5$He. In the following
we will focus only on these two isotopes since heavier hydrogen and
helium isotopes are less abundant at the conditions of relevance here
(see Fig.~\ref{fig:eos}) and play hence a minor role for the SN
dynamics and neutrino emission. Already during the final core collapse
phase, shown in Fig.~\ref{fig:fullstate_collapse}, all light clusters
are abundant. This has already been discussed in 
Refs.~\cite{Sumiyoshi:2009,Hempel:2012,Furusawa:2013} and
recently in Ref.~\cite{Yudin:2019}, however, the latter two without
performing SN simulations with light clusters explicitly taken into
account. Moreover, the larger abundance of all light clusters of the
statistical model of HS(DD2), in comparison to gRDF(DD2), is a
feedback from having lower densities with the latter SN simulations
due to the presence of heavy clusters in this density domain. Only
towards higher densities, the geometric excluded volume approach of
HS(DD2) overestimates the abundances of all light clusters, in
comparison where gRDF(DD2) already predicts their complete
dissolution. Note further that HS(DD2) does not contain any other
helium nuclei besides $^3$He and $^4$He, which results in the
significantly overestimated abundance of $^3$He. This is particularly
important since at the neutron rich conditions encountered at the
stellar core collapse, featuring $Y_e\simeq0.25-0.3$, the proton rich
nucleus $^3$He should be strongly suppressed. This is the case for
gRDF(DD2) treating in addition $^5$He which, in fact, becomes more
abundant then $^4$He under neutron rich conditions towards high
density (see Fig.~\ref{fig:fullstate_collapse}).

\subsection{Post bounce evolution}
\begin{figure*}
\subfigure[~Shock radii and neutrinospheres]
{\includegraphics[width=0.96\columnwidth]{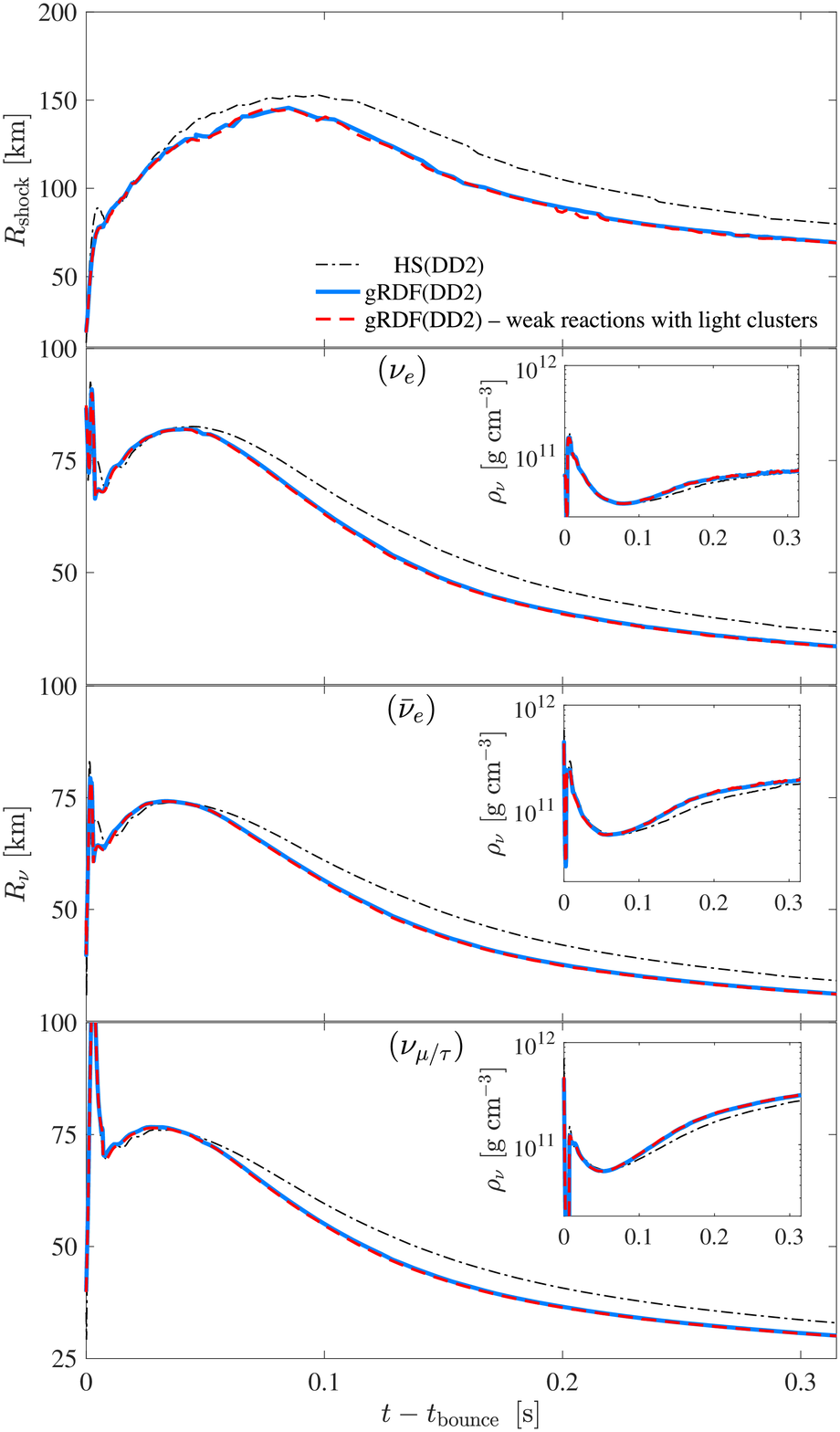}
\label{fig:shock}}
\hspace{5mm}
\subfigure[~Neutrino luminosities and average energies]
{\includegraphics[width=0.96\columnwidth]{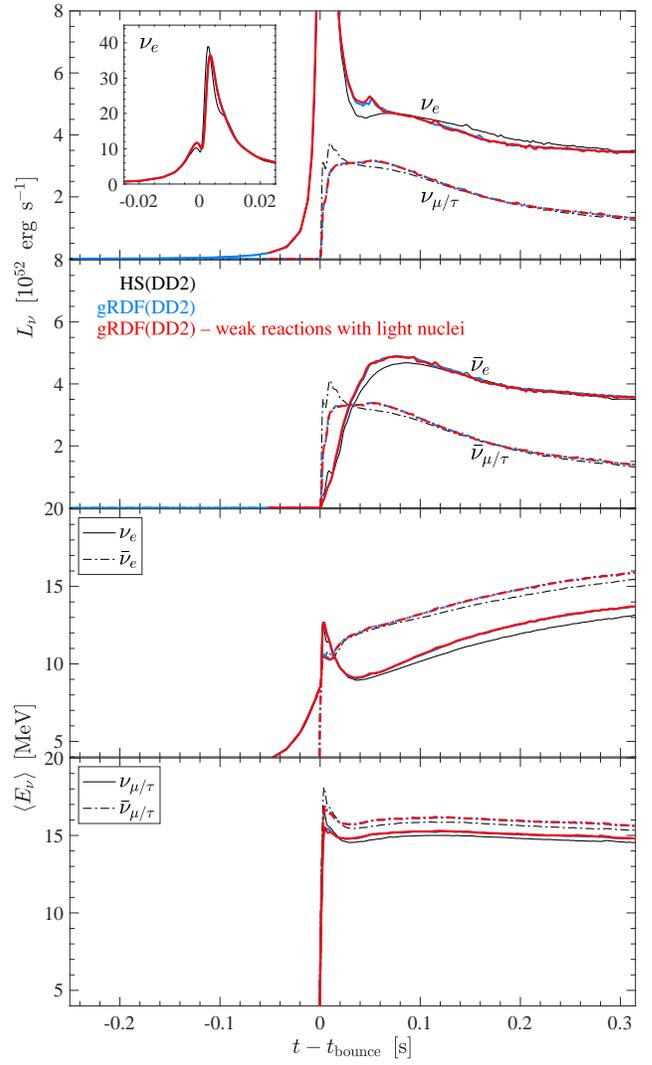}
\label{fig:neutrinos}}
\caption{Post bounce evolution of the shock radii, $R_{\rm shock}$, and neutrinosphere radii, $R_\nu$, as well as neutrino luminosity, $L_\nu$, and average energy, $\langle E_\nu \rangle$, for $\nu_e$ and $\bar\nu_e$ as well as muon- and tau-(anti)neutrinos collectively denoted as $\nu_{\mu/\tau}$ and $\bar\nu_{\mu/\tau}$, comparing the SN simulations based on the HS(DD2) EOS (black lines) and the generalized relativistic density-functional approach gDRF(DD2) without (blue lines) and with weak processes involving light clusters (red lines). Note that the latter two lay indistinguishably on top of each other (see text for details).}
\label{fig:bulk_evol}
\end{figure*}

During the early post bounce evolution, the situation remains the same as before core bounce and the differences reported previously between HS(DD2) and gRDF(DD2) still hold, as illustrated in Figs.~\ref{fig:fullstate_001pb} and \ref{fig:fullstate_010pb} at two selected post bounce times. Only when the temperature starts to exceed $T\simeq 15$~MeV the heavy nuclear clusters of gRDF(DD2) dissolve, transition into the state of homogeneous nuclear matter at around 100~ms post bounce and the SN evolution using HS(DD2) and gRDF(DD2) become increasingly similar  (see therefore Fig.~\ref{fig:fullstate_010pb} showing radial profiles of selected quantities at 100~ms post bounce). However, the higher temperatures obtained for the gRDF(DD2) EOS reflect the systematically softer EOS during the later SN post bounce evolution, beyond 100~ms. This has two important consequences, higher temperatures enable (1)~higher abundances of light clusters, as illustrated at 300~ms post bounce in Fig.~\ref{fig:fullstate_030pb}, and (2)~a more compact PNS which in turn results in a smaller shock radius already before shock stalling (see the top panel in  Fig.~\ref{fig:shock}). In particular, the faster shock retreat results in a faster PNS compression, which in turn supports even higher temperatures at the PNS interior (see Fig.~\ref{fig:fullstate_030pb}).

It it important to note further the presence of $^5$He -- a nuclear
cluster which has not been considered before in simulations of
core collapse SNe -- in the neutron rich PNS interior. The abundance
of $^5$He exceeds even the abundance of $^4$He by more than one order
of magnitude, due to the generally neutron-rich conditions featuring
$Y_e\simeq0.1-0.25$, as illustrated in the sequences shown in the
Figs.~\ref{fig:fullstate_010pb}–\ref{fig:fullstate_030pb}. Similarly,
$^4$H is found to be as abundant as $^3$H in the region of their
highest population. This is partly supported due to the neutron-rich
conditions, given by the electron fraction of $Y_e\simeq 0.1-0.2$ (see
Fig.~\ref{fig:fullstate_030pb}). The presence of a large amount of
light clusters at high densities, in the range of
$\rho=10^{12}-10^{14}$~g~cm$^{-3}$, softens the nuclear EOS in this
regime substantially, which in turn results in systematically higher
temperatures (see
Figs.~\ref{fig:fullstate_001pb}–\ref{fig:fullstate_030pb}).

Note further the systematically higher abundance of heavy clusters and
lower abundance of $^4$He in the post-shock layer ahead of the SN
accretion shock, corresponding to low densities
$\rho<10^{10}$~g~cm$^{-3}$ as well as low temperatures $T<2.5$~MeV and
nearly isospin symmetric conditions with $Y_e\simeq0.45-0.5$ (see
Figs.~\ref{fig:fullstate_bounce}--\ref{fig:fullstate_030pb}). This is
the region where the Coulomb shifts enhance the nuclear binding energy
of the heavy clusters, as discussed above in Sec.~\ref{sec:eos}.

\subsection{Neutrino emission}
The softer gRDF(DD2) EOS and the associated faster shock retreat result in an increase of the luminosities of all neutrino flavors during the early post bounce evolution up to about 100~ms, as illustrated in Fig.~\ref{fig:neutrinos}. The impact is largest for $\bar\nu_e$ and the heavy-lepton flavor neutrinos, which decouple at highest densities where the impact from the more compact PNS is largest, compared to the $\nu_e$. However, during the later post bounce phase $\gtrsim$200--250~ms, the luminosities of gRDF(DD2) match those obtained for the simulation using HS(DD2). 

The three bottom panels of Fig.~\ref{fig:shock} show the evolution of the $\nu_e$, $\bar\nu_e$ and $\nu_{\mu/\tau}$ neutrinosphere radii, as well as the corresponding densities denoted as $\rho_\nu$ in the inlays. The steeper density gradient at the PNS surface obtained for the SN simulation with the softer gRDF(DD2) EOS, results in a shift of all neutrinospheres towards smaller radii already after about 50--75~ms post bounce. This is a direct feedback from the softer gRDF(DD2) EOS and the subsequent faster PNS contraction. Complementary, the density at the neutrinospheres is shifted to slightly higher values, comparing HS(DD2) and gRDF(DD2), as shown in the inlays of Fig.~\ref{fig:shock}, featuring a slightly higher temperature. The latter aspect gives rise to generally higher average neutrino energies for all flavors for the SN simulation based on the softer gRDF(DD2) EOS, as illustrated in the bottom panels of Fig.~\ref{fig:neutrinos}. The largest increase of the average energies is obtained also for $\nu_e$ and $\bar\nu_e$, which decouple at lowest density in the layer of accumulated material at the PNS surface where the impact from the more compact PNS is largest for the gRDF(DD2) EOS, than for the heavy lepton neutrino flavors denoted as $\nu_{\mu/\tau}$ and $\bar\nu_{\mu/\tau}$ in Fig.~\ref{fig:shock} and \ref{fig:neutrinos}.

These generally higher neutrino fluxes and average energies during the long-term post bounce evolution are well known features for SN simulations employing soft EOS. However, here we are able to relate the softness of gRDF(DD2) to the presence of a large fraction of light isospin asymmetric nuclear clusters which are either neglected previously or taken incorrectly into account based on the NSE approach.

\begin{table}
\centering
\caption{Neutrino reactions with light nuclei considered, including references for the corresponding cross sections.}
\begin{tabular}{ccc}
\hline
 & Weak process & Reference \\
\hline
\hline
1 & $\nu_e + {^2}\text{H}\rightleftarrows p + p + e^-$ & \cite{Nakamura:2001} \\
2 & $\bar\nu_e + {^2}\text{H}\rightleftarrows n + n + e^+$ & \cite{Nakamura:2001} \\
3 & $\nu_e + n + n \rightleftarrows \, {^2}\text{H} + e^-$ & \cite{Nakamura:2001,Fischer:2016EPJWC,Fischer:2017} \\
4 & $\bar\nu_e + p + p \rightleftarrows \, {^2}\text{H} + e^+$ & \cite{Nakamura:2001,Fischer:2016EPJWC,Fischer:2017} \\
%5 & $\nu_e\,^3\text{H}\rightleftarrows n\,p\,p\,e^-$ \\
%6 & $\bar\nu_e\,^3\text{H}\rightleftarrows n\,n\,n\,e^+$ \\
5 & $\nu_e + {^3}\text{H}\rightleftarrows \,{^3}\text{He} + e^-$ & \cite{Fischer:2016EPJWC,Fischer:2017} \\
6 & $\bar\nu_e + {^3}\text{He}\rightleftarrows \,{^3}\text{H} + e^+$ & \cite{Fischer:2016EPJWC,Fischer:2017} \\
7 & $\nu_e + {^4}\text{H}\rightleftarrows \,{^4}\text{He} + e^-$ & this work \\
8 & $\bar\nu_e + {^4}\text{He}\rightleftarrows \,{^4}\text{H} + e^+$ & this work \\
\hline
\end{tabular}
\label{tab:nu-reactions-light}
\end{table}
%

%%%%%%%%%%%%%%%%%%%%%%%%%%%%%%%%%%%%%%%%%%%%%%%%%%%%%%%%%%%%%%%%%%%%%%%%%%%%%%%%%%%%%%%%
\section{Role of weak reactions with light clusters}
\label{sec:v-reactions}
Then presence of light nuclear clusters enables a variety of weak processes. The ones considered here are listed in Table~\ref{tab:nu-reactions-light}. These concern the charged current neutrino emission and absorption of $\nu_e$ and $\bar\nu_e$ on light clusters. Neutral current neutrino scattering processes involving light clusters are omitted here included in the collision integral of the Boltzmann equation based on the coherent scattering formalism of Ref.~\cite{Bruenn:1985en}. However, the contribution of coherent scattering on all neutron rich light clusters to the inverse neutrino mean-free path is negligible~\cite{Fischer:2017}. Instead, this channel is dominated by  neutrino-nucleon scattering above scattering on light clusters, by several orders of magnitude in the high density region where these light clusters are most abundant (see Figs.~\ref{fig:fullstate_001pb}--\ref{fig:fullstate_030pb}). Hence, here we focus on charged current absorption processes. 

\begin{table*}
\centering
\caption{Two thermodynamic conditions (A) and (B), as well as the corresponding mass fractions, $X_i$, and mean-field potentials, $U_i$, for the unbound nucleons and the light clusters.}
\begin{tabular}{c|c|c|c|c|c|c|c|c|c|c|c|c|c|c|c|c|c}
\hline
\hline
 & $T$ & $\rho$ & $Y_e$ & $X_n$ & $X_p$ & $X_{^2 \rm H}$ & $X_{^3 \rm H}$ & $X_{^4 \rm H}$  & $X_{^3 \rm He}$ & $X_{^4 \rm He}$ & $U_n$ & $U_p$ & $U_{^2 \rm H}$ & $U_{^3 \rm H}$ & $U_{^4 \rm H}$ & $U_{^3 \rm He}$ & $U_{^4 \rm He}$ \\
& $[$MeV$]$ & $[$g~cm$^{-3}]$ & &  &  &  &  &  &  &  & $[$MeV$]$ & $[$MeV$]$ & $[$MeV$]$ & $[$MeV$]$ & $[$MeV$]$ & $[$MeV$]$ & $[$MeV$]$ \\
\hline
(A) & 5 & $5\times 10^{12}$ & 0.2 & 0.5 & 0.025 & 0.073 & 0.069 & 0.075  & 0.004 & 0.036 & $-1.5$ & $-3.3$ & $-3.0$ & $-0.85$ & $-1.9$ & $-4.9$ & $-2.3$ \\
(B) & 7 & $2\times 10^{13}$ & 0.05 & 0.8 & 0.001 & 0.004 & 0.003 & 0.0014 & 0.0003 & 0.0009 & $-6.3$ & $-4.5$ & $-10.02$ & 0.77 & 14.7 & $-25.0$ & $-15.2$ \\
\hline
\end{tabular}
\label{tab:conditions_II}
\end{table*}
\begin{figure*}
\includegraphics[width=0.98\columnwidth]{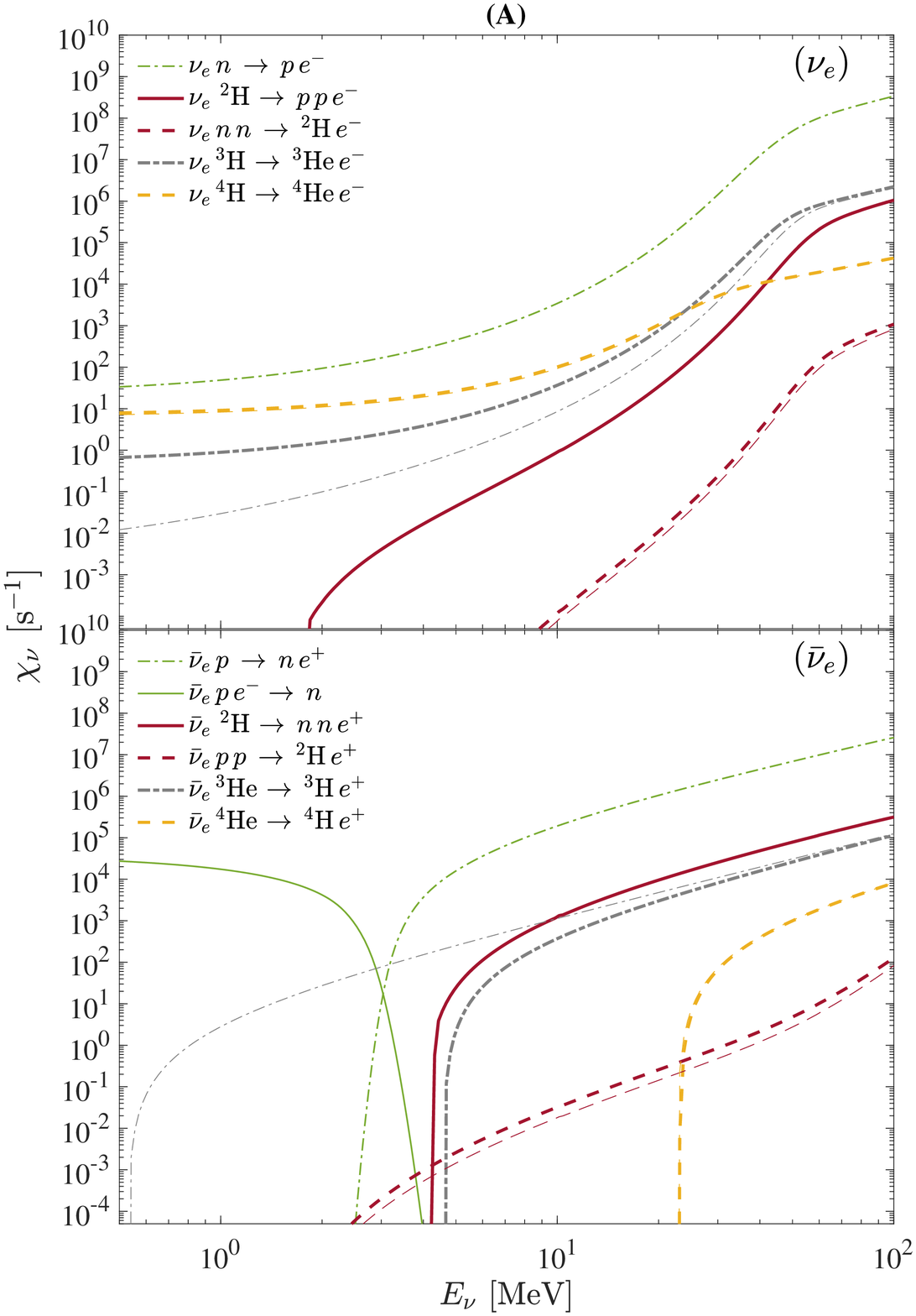}
\hspace{5mm}
\includegraphics[width=0.98\columnwidth]{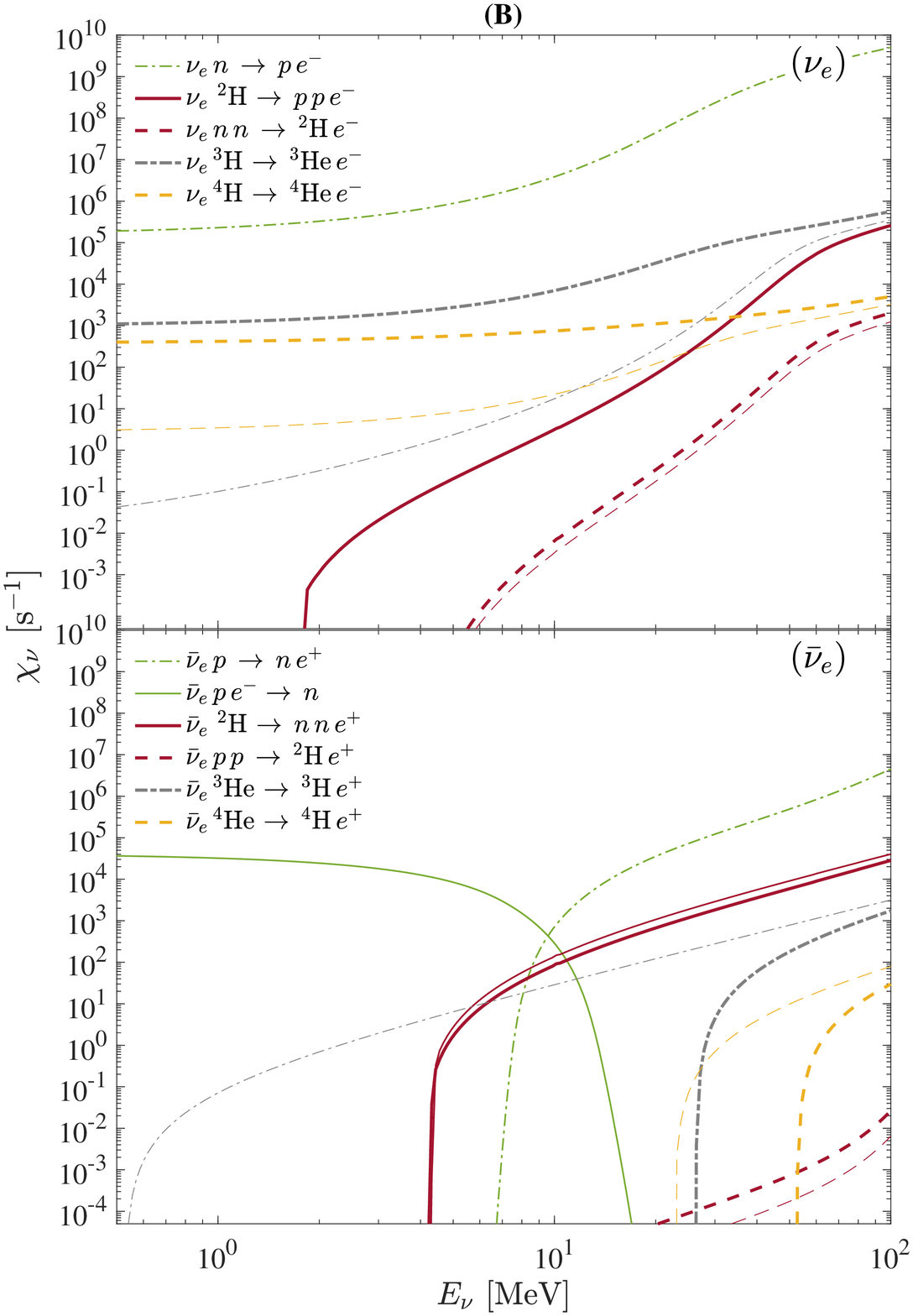}
\caption{Charged current neutrino (top panels) and antineutrino opacity (bottom panels) 
comparing the processes with the unbound nucleons (green dash-dotted lines) including 
the inverse neutron decay (green solid lines) for $\bar\nu_e$, and (anti)neutrino 
absorption on $^2$H (solid dark red lines) as well as the inverse $e^\pm$ captures 
on $^2$H (dark red dashed lines), (anti)neutrino absorption on $^3$H and $^3$He 
(grey dash-dotted lines), and (anti)neutrino absorption on $^4$H and $^4$He 
(yellow dashed lines), for the two selected conditions (A) and (B) listed in 
Table~\ref{tab:conditions_II} in the left and right panels, respectively. Thick 
lines include all medium modifications, see therefore the rate expressions in 
Appendix~\ref{sec:Appendix_c}, while thin lines omit them.}
\label{fig:rates}
\end{figure*}

\subsection{Charged current weak rates with light clusters}

The reaction rates involving light clusters with $1<Z<2$ and $2<A<4$ are provided in  Appendix~\ref{sec:Appendix_c1}--\ref{sec:Appendix_c4}. In the following the opacity 
expressions for $\nu_e$ and $\bar\nu_e$ absorption on $^2$H~\eqref{eq:cc_ve_deuteron}, 
inverse electron/positron captures on $^2$H~\eqref{eq:cc_ec_deuteron}, $\nu_e$ 
absorption on $^3$H~\eqref{eq:cc_ve_triton}, $\nu_e$ absorption on $^3$He~\eqref{eq:cc_veb_triton} are being evaluated at two representative conditions 
(A) and (B) listed in Table~\ref{tab:conditions_II}. The corresponding thermodynamic 
state is lso given in Table~\ref{tab:conditions_II}, listing the mass fractions for the 
unbound nucleons and for the light clusters as well as their mean-field potentials. 
The corresponding weak rates involving these light nuclei are shown in Fig.~\ref{fig:rates} 
for the selected conditions (A) and (B). Besides the standard charged current reactions 
involving the unbound nucleons, for the processes involving $^2$H the rate expressions 
are derived in Appendix~\ref{sec:Appendix_c1} and Appendix~\ref{sec:Appendix_c2} 
for the cross sections provided in Ref.~\cite{Nakamura:2001}, see also 
Ref.~\cite{Shen2012}. For the remaining momentum and angular integrals a 
64-point and 32-point Gauss-Legendre numerical integration scheme is implemented, 
respectively. For the reactions involving $^3$H and $^3$He the bound state to 
bound state transitions are employed here (see Appendix~\ref{sec:Appendix_c3}), 
which are favored, due to the small $Q$-value of $m_{^3 \rm H}-m_{^3 \rm He}=0.529$~MeV
(including electron contributions), above the spallation reactions. The latter processes,
transition into the continuum with three final state nucleons, are strongly suppressed 
due to the large $Q$-values, 8.44~MeV (for $\nu_e$) and 9.7~MeV (for $\bar\nu_e$), 
besides the 3-body kinematics which further reduces the rates. Furthermore, the corresponding 
matrix element for the reactions involving $^3$H and $^3$He is well known from the 
triton decay.

\begin{figure*}
\centering
\subfigure[~Mass fractions for selected nuclear species]{\includegraphics[width=1.8\columnwidth]{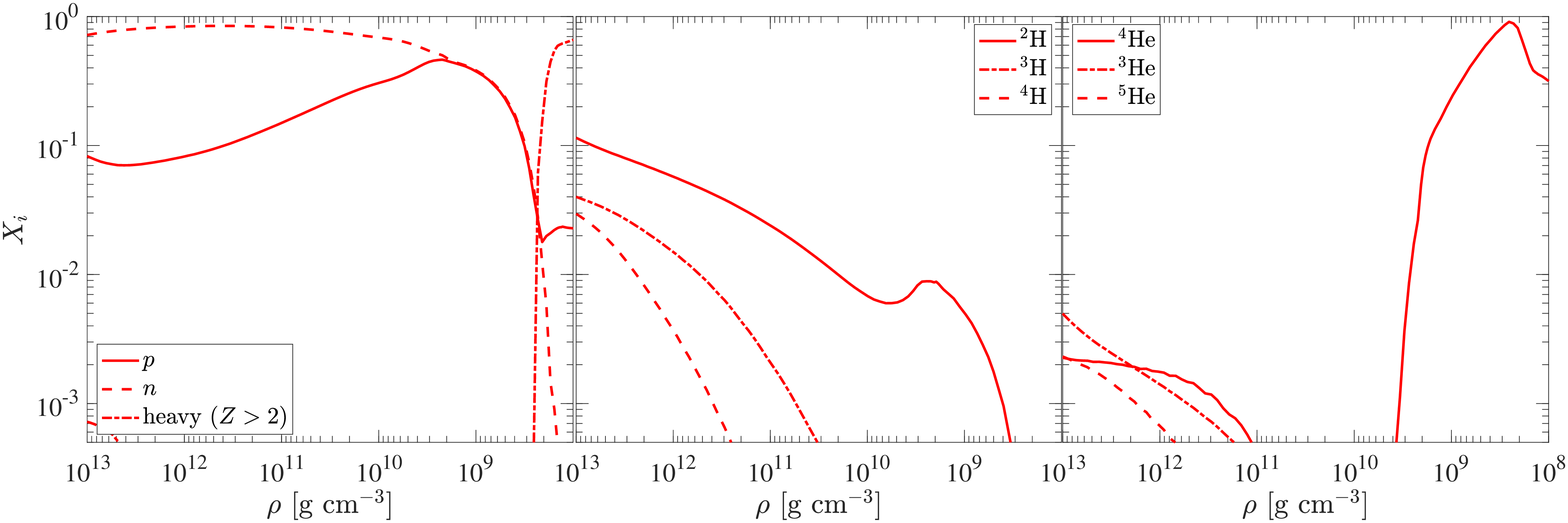}\label{fig:abund}}
\\
\subfigure[~Net neutrino heating rates (top panels) and inverse mean-free paths for selected weak reactions (bottom panels)]{\includegraphics[width=1.8\columnwidth]{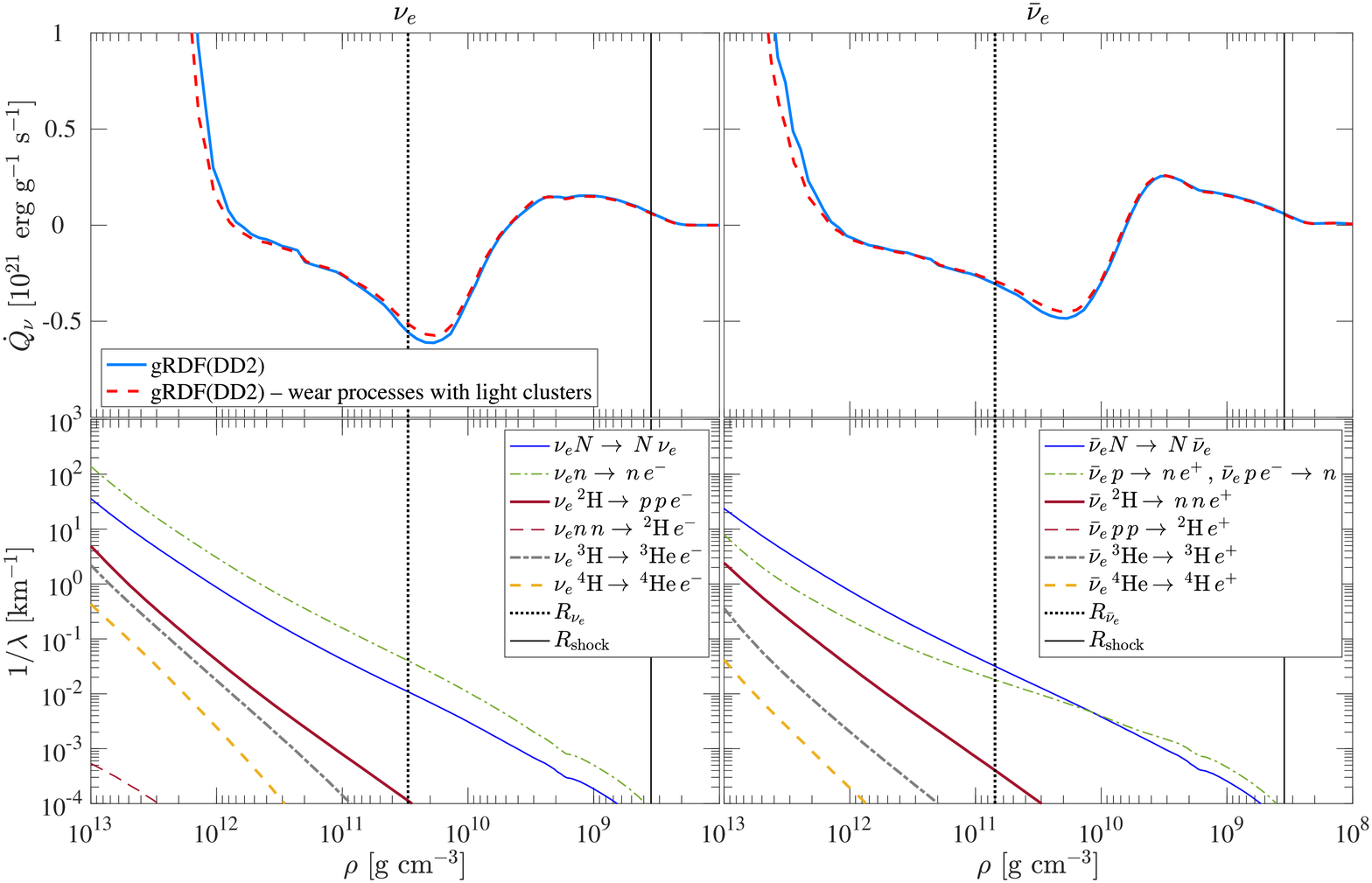}\label{fig:mfp}}
\caption{SN simulation profiles with respect to the restmass density at 100~ms post bounce. Graph~(a): mass fractions, $X_i$, for selected nuclear species. Graph~(b): net neutrino heating rate, $\dot Q_\nu$, comparing gRDF(DD2) with (red dashed lines) and without (solid light blue lines) weak reactions involving light nuclei for $\nu_e$ (left panel) and $\bar\nu_e$ (right panel), as well as the inverse neutrino mean-free path, $1/\lambda$, for selected weak processes corresponding to the simulations with the gRDF(DD2) EOS including weak reactions with light clusters, also for $\nu_e$ (left panel) and $\bar\nu_e$ (right panel), distinguishing neutral current neutrino-nucleon scattering (solid blue lines), charged current neutrino absorption with unbound nucleons (dark greed dash-dotted lines), $^2$H (thick solid dark red lines), inverse $e^\pm$ captures on $^2$H (thin dashed dark red lines), involving $^3$H and $^3$He (thick grey dash-dotted lines) as well as $^4$H and $^4$He (thick dashed dark yellow lines). The vertical black lines mark the locations of the neutrinospheres (dotted lines) and the shock position (solid lines).}
\end{figure*}

In addition we also consider charged current processes with $^4$H and $^4$He (see Appendix~\ref{sec:Appendix_c4}), for which also only the bound state to bound state transitions are employed with a large $Q$-value, $m_{^4 \rm H}-m_{^4 \rm He}=22.7$~MeV, which is modified due to the mean-field potential differences. The latter contributions, $U_{^4 \rm H}-U_{^4 \rm He}$, can be large for these reactions at the conditions of interest here (see Table~\ref{tab:conditions_II}). Furthermore, $^4$H is unstable under laboratory conditions, it decays via the emission of a neutron to triton on the strong-interaction timescale. However, the presence of the nuclear medium enables the following strong equilibrium, $^3 \rm H + n \leftrightarrows\,^4$H, which yields a finite abundance of $^4$H. The same applies to all other neutron rich hydrogen and helium isotopes which are also unstable due to neutron emission under vacuum conditions, except $^6$He which beta-decays to $^6$Li. Details about the properties of $^4$H, as well as all other neutron-rich hydrogen and helium isotopes, and how to implement medium modifications was discussed above in Sec.~\ref{sec:eos}. Important for the neutrino response involving these nuclei is the fact that a weak interaction matrix element for the charged current transitions is presently unknown, both experimentally for the obvious reason but also theoretically. In order to study their potential role on the neutrino emission as well as heating and cooling in SN simulations, we assume a Gammow-Teller transition amplitude. Similar as for the reactions involving $^3$H and $^3$He for which the transition amplitude is known from the triton decay, for reactions involving $^4$H and $^4$He we derive the weak transition amplitude from the thermally averaged cross sections for $\bar\nu_e$ absorption on $^4$He of Ref.~\cite{Gazit:2007PhRvL}. Details, including the cross sections, are given in Appendix~\ref{sec:Appendix_c4}.

Figure~\ref{fig:rates} compares the charged current opacity, $\chi_\nu$, for all charged current weak processes listed in Table~\ref{tab:nu-reactions-light} involving light nuclei with those of the unbound nucleons, for $\nu_e$ (top panels) and $\bar\nu_e$ (bottom panels), at two selected conditions (A) and (B) listed in Table~\ref{tab:conditions_II}. They correspond roughly to regions in a core collapse SN simulation where light clusters are most abundant, in particular where the light cluster abundances exceed the abundance of protons, as listed in Table~\ref{tab:conditions_II}. Here it becomes evident that at any of these conditions the reaction rates with the unbound nucleons largely exceed those of the light clusters (see Fig.~\ref{fig:rates}). Note the opacity dropping to zero, which corresponds roughly to the medium-modified $Q$-values which are a sharp threshold for the reactions treated within the elastic approximation, i.e. involving $^3$H and $^3$He as well as $^4$H and $^4$He. The mean-field potentials for these nuclei are given in Table~\ref{tab:conditions_II}. They are provided by the gRDF(DD2) EOS. 

\subsection{SN simulations with weak reactions involving light clusters}
All charged current weak reactions listed in Table~\ref{tab:nu-reactions-light} are implemented into the transport module of {\tt AGILE-BOLTZTRAN}, based on the explicit rate expressions for the neutrino emissivity and opacity as given in Appendix~\ref{sec:Appendix_c1}--\ref{sec:Appendix_c5}. We use the same neutrino transport discretization as before. The core collapse SN simulations employ then the new gRDF(DD2) EOS, which provide now also the information about the mean-filed potentials for all light clusters. These enter explicitly all rate expressions (see Appendix~\ref{sec:Appendix_c1}--\ref{sec:Appendix_c5}). 

The bulk evolution, from the shock formation up to several hundreds of ms post bounce,
proceeds indistinguishable from the reference setup without considering weak reactions
with light nuclei, as shown in Fig.~\ref{fig:shock}. The impact from the additional weak reactions
involving light clusters to the neutrino heating and cooling is generally small, as illustrated
in Fig.~\ref{fig:mfp} at 100~ms post bounce. In particular the contribution to the neutrino
heating, which was suggested previously \cite{Furusawa:2013} where neutrino emission 
contributions were neglected which violates the relation of detailed balance, could not be 
found. Instead, the inclusion of weak reactions with light clusters leaves a negligible impact
on the overall post bounce evolution, as illustrated in the evolution of the neutrinospheres
in Fig.~\ref{fig:shock} as well as the neutrino luminosities and average energies in 
Fig.~\ref{fig:neutrinos}, in comparison to the simulations with gRDF(DD2) omitting weak
reactions with light clusters.

This negligible impact on the post bounce evolution can be understood through the analysis of the neutrino heating/cooling rates, shown in the top panels of Fig.~\ref{fig:mfp} at about 100~ms post bounce, for $\nu_e$ (left panel) and $\bar\nu_e$ (right panel). The comparison between the two SN simulations employing the gRDF(DD2) EOS without (solid blue lines) and with weak reactions involving light clusters (dashed red lines) reveals only minor differences. These are attributed not to additional heating and cooling contributions stemming from weak reactions with light clusters, but rather a slight mismatch of the timing between these two simulations on the order of 10~ms. Note the steep rise of the heating rates towards higher densities, which is due to the fact that the {\tt AGILE-BOLTZTRAN} finite differencing scheme is not specially tuned to enforce the diffusion limit, for which the net rates would be zero (details can be found in Ref.~\cite{Liebendoerfer:2004}).

At the conditions where the neutrinos decouple, see therefore the locations of the accretion shock (vertical solid black lines) and of the neutrinospheres (vertical dashed black lines), the abundances of all neutron-rich light clusters are negligible in comparison to the abundances of the unbound nucleons. Consequently, the inverse mean-free path, shown in the bottom panels of Fig.~\ref{fig:mfp}, for $\nu_e$ (left panels) and $\bar\nu_e$ (right panels), are dominated by weak reactions involving the unbound nucleons. Note that the definition of the inverse mean-free path is given in Ref.~\cite{Fischer:2012a}, for which the neutrino phase-space integration is performed numerically using the neutrino distributions obtained from the SN simulations, following our previous works~\cite{MartinezPinedo:2014,Fischer:2020}. In addition to the charged current absorption reactions for $\nu_e$ and $\bar\nu_e$ (green dash-dotted lines), we also include the elastic neutrino-nucleon scattering processes (solid blue lines). The latter dominates the inverse mean-free path for $\bar\nu_e$. Note that for $\bar\nu_e$ the inverse mean-free path for the inverse electron captures on $^2$H (dashed red line) is below the scale of the plot. The situation remains the same also at later times. 

Only towards higher density, in excess of few times $10^{13}$~g~cm$^{-3}$, there are contributions from weak interactions with light clusters, in particular involving $^2$H. This is related with the increasing abundances of all light clusters towards higher densities, as illustrated and discusses above in Sec.~\ref{sec:SN-sim}. However, firstly, their contribution remains negligible and secondly, at these high densities the entire neutrino spectrum is trapped such that it leaves no impact on the post bounce evolution. The situation may change when the neutrinospheres shift to higher densities during the PNS deleptonization after the SN explosion onset has been launched and the nascent PNS deleptonizes and cools via the emission of neutrinos of all flavors, which remains to be explored in the future.

\section{Summary}\label{sec:summary}
The present article introduces the gRDF(DD2) EOS for astrophysical simulations. 
Particular focus is devoted to the improved description of light and heavy nuclear 
clusters, taking consistently into account medium modifications of the nuclear binding 
energies with contributions due to Pauli blocking and Coulomb corrections. The
gRDF(DD2) EOS includes self consistently neutron-rich hydrogen ($A=2-7$) and
helium isotopes ($A=3-8$), some of which are weakly bound such as the hydrogen 
isotopes with $A>3$ as well as $^5$He, and were hence omitted previously. Their
role was recently overestimated due to the implementation of the NSE approach
\cite{Yudin:2019} which is based on vacuum binding energies. This caveat has
been overcome within the present work taking explicitly into account individual 
scattering state contributions following the QS approach of Ref.~\cite{Roepke:2020},
which solves the in-medium Schr{\"o}dinger equation. We find that the continuum
contributions, known as virial corrections, provide the essential contributions to all
weakly bound clusters such as $^4$H and $^5$He, while they play only a minor
role for the bound clusters such as $^3$H, $^3$He and $^4$He. For $^2$H the 
virial corrections are known to have contributions already at low densities 
\cite{OConnor:2007,Horowitz:2006}.%, which results in a reduction of the deuteron density by a factor of about 0.54 in comparison to NSE. 

While the QS framework allows for the calculations of individual light clusters,
it cannot be applied for astrophysical EOS which require the inclusion of more
than 1000 nuclear species, in particular also heavy nuclei. Therefore, the gRDF
approach has been developed here, which is based on the nuclear in-medium 
binding energies provided for selected light nuclear clusters from the QS calculations. 
Consequently, gRDF results in good quantitative agreement with the quantum-statistical
approach in terms of the properties of the light nuclear clusters. Deviations are 
observed only for  the weakly bound light clusters. In particular, both approaches 
gRDF and QS predict the dissolving of the light nuclear clusters at substantially lower
density, already at about one tenth of the nuclear saturation density, than the NSE 
approach of Ref.~\cite{Hempel:2009mc}, which implements an excluded volume 
mechanism instead. A qualitative comparison of the NSE based excluded volume 
approach and the QS can be found in Ref.~\cite{Hempel:2011}. Here, gRDF provides 
the general description for all clusters where the heavy clusters, defined as $Z>2$, 
are taken into account following Ref.~\cite{Pais:2016fdp}. This results in the presence 
of heavy clusters up to high densities, on the order of the saturation density, and 
hence a late transition to homogeneous matter. 

These aspect of the description of light and heavy clusters significantly improve the 
commonly employed approaches, not only the excluded volume of Ref.~\cite{Hempel:2009mc} 
based on NSE, but also the single-nucleus approximation employed 
\cite{Lattimer:1991nc,Furusawa:2013b,Schneider:2017} featuring a phase 
transition construction for the transition to homogeneous matter, the Thomas-Fermi 
approximation \cite{Shen:1998by,Pais:2018} and the spherical Wigner-Seitz cell approach
\cite{gshen2011b}.

We implement the new gRDF(DD2) EOS into our spherically symmetric core collapse 
SN model {\tt AGILE-BOLTZTRAN} and perform SN simulations, launched from the 
18~M$_\odot$ progenitor \cite{Woosley:2002zz}. In comparison with the HS(DD2) 
NSE EOS we find a substantially different composition, particularly significantly lower
abundances of all light clusters and the presence of $^4$H and $^5$He which were 
omitted in previous SN simulations. Furthermore, towards high densities at the PNS 
interior, the abundances of $^5$He exceed those of $^3$He and $^4$He due to the 
generally neutron-rich conditions with low $Y_e=0.1-0.25$. This is also the reason 
for the high yields of $^3$H and $^4$H, being comparable to those of unbound protons. 
Furthermore, the heavy cluster abundance at high densities has a strong softening 
impact on the PNS structure, already around the stellar core bounce, which in turn 
results in lower central densities and higher temperatures. Also the emission of the 
$\nu_e$ deleptonization burst, associated with the bounce shock propagation across 
the $\nu_e$ sphere, is affected from the presence of more heavy clusters, featuring 
a lower abundance of unbound nucleons and hence a slight reduction of the magnitude 
as well as a miner broadening. Furthermore, the generally softer high-density behaviour 
of the gRDF(DD2) EOS results in a faster shock retreat and a smaller shock radius 
during the post bounce evolution up to several hundreds of milliseconds. It affects 
the post bounce neutrino emission, with generally higher average neutrino energies. 

The spherically symmetric simulations presented here, featuring a faster shock retreat 
due to the softer gRDF(DD2) EOS, in comparison to the HS(DD2) EOS, result in less 
favourable conditions for shock revival in terms of neutrino heating. However, similar 
comparisons in the multi-dimensional SN framework indicate the opposite effect with 
respect to shock revival via neutrino heating (c.f. Ref.~\cite{Suwa:2013}). This remains 
to be explored in multi-dimensional SN simulations.

The gRDF(DD2) EOS provides not only the gross thermodynamics properties of 
heavy and light nuclear clusters, but also the microscopic quantities such as scalar 
and vector self energies for the unbound nucleons and for all light clusters. The 
latter enables us to consistently treat a variety of weak processes with light clusters, 
which were either omitted in previous core collapse SN studies or crudely approximated. 
While it has been shown previously that neutral current scattering reactions involving 
light clusters are negligible, instead this channel is dominated by neutrino scattering 
off unbound nucleons, here we focus on the charged current absorption reactions. 
We introduce the explicit rate expressions for the spallation reactions for $^2$H as 
well as electron/positron captures on $^2$H, based on the cross sections of 
Ref.~\cite{Nakamura:2001}. Further reaction rates involving $^3$H and $^3$He 
as well as $^4$H and $^4$He are provided based on the elastic approximation 
where we consider the bound state to bound state transition since the transitions 
into the continuum are negligible \cite{Fischer:2017}. These charged current weak 
rates are implemented into the collision integral of {\tt AGILE-BOLTZTRAN} and 
additional SN simulations are performed. The comparison with the gRDF(DD2) 
EOS reference case, where these weak processes were omitted, show a negligible 
impact from the inclusion of weak reactions with these light clusters. Nevertheless, 
we would like to stress the demand for more sophisticated calculations of 
charged current neutrino absorption cross sections for the neutron-rich nuclei 
introduced here, in particular taking into account the broad width of all neutron-rich 
hydrogen isotopes, nuclear structure effects for the vacuum-stable nuclei, e.g., 
$^4$He and $^6$He as well as generally final-state and medium contributions.

\section*{Acknowledgement}
T.F. and N.U.F.B. acknowledge support from the Polish National Science Center (NCN) under the following grant numbers 2016/23/B/ST2/00720 (T.F.), 2019/33/B/ST9/03059 (T.F.) and 2019/32/C/ST2/00556 (N.U.F.B.). The supernova simulations are performed at the Wroclaw Center for Scientific Computing and Networking (WCSS) in Wroclaw (Poland).

\appendix

\section{gRDF Relativistic mean-field model revisited}\label{sec:Appendix_gRDF}
All EOS quantities that are relevant for SN simulations can be derived from the grand canonical potential density, $\omega$, of the gRDF model. It depends on the temperature, $T$, and the set of chemical potentials, $\{\mu_{i}\}$, of all considered particles. In this formulation chosen here, meson fields of the RMF model do not appear explicitly and only baryons are considered as effective degrees of freedom. Then the grand canonical potential density can be written as follows:
\begin{equation}
\label{eq:omega}
\omega(T,\{\mu_{i}\}) = \sum_{i\in \mathcal{B}} \omega_{i} + \omega_{\rm meson} - \omega^{(r)}~,
\end{equation}
with contributions from baryons, $\mathcal{B} = \left\{ n,p,{}^{2}\mbox{H}, {}^{3}\mbox{H}, \dots \right\}$, i.e. nucleons and nuclei, $\omega_i$, as well as mesons $\omega_{\rm meson}$ and rearrangement terms denoted as $\omega^{(r)}$.  The baryons are treated as quasi-particles with effective chemical potentials $\mu_{i}^{\ast} = \mu_{i}-
\Sigma^V_{i}$ and with the dispersion relation,
\begin{equation}
E_{i}(p,m_{i}^{\ast}) = \sqrt{p^{2}+(m_{i}^{\ast})^{2}}
\end{equation}
depending on the momentum $p$ and the effective mass $m_{i}^{\ast} = m_{i}-\Sigma^S_{i}$~. The quantities $\Sigma^V_{i}$ and $\Sigma^S_{i}$ are the vector and scalar self-energies or potentials of the baryons, see below, and $m_{i}$ is the rest mass in vacuum. The particle number and scalar densities are defined by the standard expressions
\begin{equation}
n_{i} = - \left. \frac{\partial \omega}{\partial \mu_{i}} \right|_{T,\{ \mu_{j\neq i}\}}~,
\qquad
n_{i}^{(s)} = \left. \frac{\partial \omega}{\partial m_{i}} \right|_{T,\{ \mu_{j}\}}~
\end{equation}
as derivatives of $\omega$ with respect to the chemical potentials $\mu_{i}$ and masses $m_{i}$, respectively.
The contribution of baryons to the total grand canonical potential density \eqref{eq:omega} has the form
\begin{widetext}
\begin{equation}
\label{eq:omega_i}
\omega_{i}^{(0)}(T,m_{i}^{\ast},\mu_{i}^{\ast}) = - T \frac{g_{i}(T)}{\sigma_i}\,\int \frac{d^{3}p}{(2\pi)^{3}}\,\ln\left\{ 1 + \sigma_i \exp\left[ -\frac{E_{N}(p,m_{i}^{\ast}) -\mu_{i}^{\ast}}{T}\right]\right\}~,
\end{equation}
\end{widetext}
with particle degeneracy factors $g_{i}$ that depend on temperature in the case of heavy nuclei in order to take the contribution of excited states into account. The quantity $\sigma_{i}$ encodes the particle statistics (Fermi-Dirac: $\sigma_i=+1$, Bose-Einstein: $\sigma_i=-1$). For heavy nuclei, the non-relativistic limit of the energy, $E_{i} = m_{i}+p^{2}/(2m_{i})$, and Maxwell-Boltzmann statistics are assumed, i.e., the limit $\sigma_i \to 0$ is taken in equation~\eqref{eq:omega_i}.  
The meson contribution,
\begin{equation}
\omega_{\rm meson} = - \frac{1}{2} \left( C_{\omega} n_{\omega}^{2}  + C_{\rho} n_{\rho}^{2} - C_{\sigma} n_{\sigma}^{2} \right)~,
\end{equation}
to \eqref{eq:omega} is expressed with the help of the coupling factors $C_{j}=\Gamma_{j}^{2}/m_{j}^{2}$ of the mesons $j\in \mathcal{M}=\left\{ \omega,\rho,\sigma \right\}$ and their source densities $n_{j}$. 
The couplings $\Gamma_{j}$ depend on the total baryon density ${n}_{B}=\sum_{i\in B}B_{i}n_{i}$ with baryon numbers $B_{i}$ and baryon number densities $n_{i}$. The density dependence of the couplings causes the appearance of a meson rearrangement term
\begin{equation}
V_{\rm meson}^{(r)} = \frac{1}{2} \left( C_{\omega}^{\prime} n_{\omega}^{2} + C_{\rho}^{\prime} n_{\rho}^{2} - C_{\sigma}^{\prime} n_{\sigma}^{2} \right)
\end{equation}
with derivatives $C_{j}^{\prime} = dC_{j}/dn_{B}$ in the vector potential
\begin{equation}
\Sigma^V_{i} = C_{\omega} g_{i\omega} n_{\omega} + C_{\rho} g_{i\rho} n_{\rho} + B_{i} V_{\rm meson}^{(r)} + W_{i}^{(r)}
\label{eq:app_sigmaV}
\end{equation}
besides the usual meson contributions in RMF models and a further rearrangement term, 
\begin{equation}
W_{i}^{(r)} = \sum_{j\in \mathcal{B}} n_{j}^{(s)} \frac{\partial \Delta m_{j}}{\partial n_{i}}~,
\end{equation}
due to the medium dependence of the masses. The scalar potential
\begin{equation}
\Sigma^S_{i} = C_{\sigma} g_{i\sigma} n_{\sigma} - \Delta m_{i}
\label{eq:app_sigmaS}
\end{equation}
explicitly includes the medium dependent mass shift $\Delta m_{i}$ for all baryons except nucleons.
The degeneracy factors $g_{i}$ as well as the factors $g_{ij}$ in the vector and scalar potentials are specified  Ref.~\cite{Pais:2016fdp}. The latter also appear in the source densities,
\begin{equation}
    n_{\omega} = \sum_{i\in \mathcal{B}} g_{i\omega} n_{i}~,
    \quad
    n_{\rho} = \sum_{i\in \mathcal{B}} g_{i\rho} n_{i}~,
    \quad
    n_{\sigma} = \sum_{i\in \mathcal{B}} g_{i\sigma} n_{i}^{(s)}~.
\end{equation}
The mass shifts $\Delta m_{i} = \Delta m_{i}^{\rm (Coul)} + \Delta m_{i}^{\rm (strong)}$ in the scalar potential contain a Coulomb contribution due to the screening of the Coulomb field of the nuclei by the charged leptons and a strong shift that arises from the action of the Pauli principle. It causes a blocking of states in the medium that are no longer available for the formation of nuclei as many-body correlations. The mass shifts depend on the particle densities and the temperature. Explicit expressions are given Ref.~\cite{Pais:2016fdp}. Finally, the last term in \eqref{eq:omega} has the form
\begin{equation}
\omega^{(r)} = V_{\rm meson}^{(r)} n_{B} + \sum_{i\in \mathcal{B}} W_{i}^{(r)} n_{i}
\end{equation}
with the contributions from the density dependence of the couplings and the mass shifts. The rearrangement
terms in all quantities are essential for the thermodynamic consistency of the model and lead to the standard expressions for the densities of the quasi-particles,
\begin{equation}
n_{i} = g_{i} \int \frac{d^{3}p}{(2\pi)^{3}} \: f_{i}~,
\qquad
n_{i}^{(s)} = g_{i} \int \frac{d^{3}p}{(2\pi)^{3}} \: f_{i} \: \frac{m_{i}^{\ast}}{E_{i}}
\end{equation}
with the distribution function
\begin{equation}
f_{i}(T,p,m_{i}^{\ast},\mu_{i}^{\ast}) = \left\{ \exp \left[ \frac{E_{i}(p,m_{i}^{\ast})-\mu_{i}^{\ast}}{T}\right] + \sigma_{i} \right\}^{-1}~.
\end{equation}
The grand canonical potential density \eqref{eq:omega} immediately gives the pressure $p=-\omega$ of the system and the free energy density, $f$, is found from the thermodynamic relation
\begin{equation}
f = -p + \sum_{i\in \mathcal{B}} \mu_{i} n_{i}~.
\end{equation}
The entropy density, $s$, is obtained from the standard definition as
\begin{widetext}
\begin{equation}
    s = - \left. \frac{\partial \omega}{\partial T} \right|_{\{\mu_{i}\}} =
    - \sum_{i\in \mathcal{B}} g_{i} \int
    \frac{d^{3}p}{(2\pi)^{3}} \: \left[ f_{i} \ln f_{i} + \frac{1-\sigma_{i}f_{i}}{\sigma_{i}} 
    \ln \left(1-\sigma_{i}f_{i}\right)\right]
    - \sum_{i\in \mathcal{B}} \frac{\omega_{i}^{(0)}}{g_{i}} \frac{\partial g_{i}}{\partial T}
\end{equation}
with an contribution that originates from the temperature dependence of the degeneracy factors of nuclei.

\section{Virial expansion and generalized Beth-Uhlenbeck equation}\label{sec:Appendix_BU}
Within a  consistent QS approach, the densities of nucleons (as function of $T$ and the chemical potentials, e.g., of the neutron $\mu_n$ and the proton $\mu_p$) are decomposed into partial densities of different channels $\{A,Z\}$ which contain the bound states (ground state and excited states) of the nucleus with mass number $A$ and proton number $Z$, as well as the continuum states. As example, in the $^2$H channel we obtain the generalized Beth-Uhlenbeck equation as follows (for details, see Ref.~\cite{Roepke:1990}),
\begin{equation}
\label{gBUd}
n_{^2 \rm H}(T,\mu_n,\mu_p) = \frac{3}{(\pi \hbar^2/mT)^{3/2}}\,\exp\left(\frac{\mu_n+\mu_p}{T}\right) 
\,\left\{\left[\exp\left(-\frac{E_{^2 \rm H}}{T}\right)-1\right] + \frac{1}{\pi T} \int_0^\infty dE\, \exp\left(-\frac{E}{T}\right)\left[ \delta_{p,n}(E)-\frac{1}{2} \sin\left(2\delta_{p,n}(E)\right)\right]\right\}~.
\end{equation}
The two-nucleon bound state $E_{^2{\rm H}}$ and phase shifts $\delta_{p,n}$ are solutions of an in-medium two-nucleon Schr{\"o}dinger equation which is derived from a Green's function approach. In contrast to the simple Beth-Uhlenbeck equation where the deuteron bound state energy $E^{(0)}_{^2 \rm H} = -2.225$ MeV appears, the medium modified binding energy is taken. The phase shifts $\delta_{p,n}(E)$ are also modified by in-medium effects such as self-energy and Pauli-blocking terms. Because the single-nucleon contribution contains already the quasiparticle shift, e.g. within the gRDF(DD2) approach, the sin-term in the continuum contribution appears to avoid double counting.

Calculations show that the simple ground-state contribution to the partial density is reduced if the continuum contributions are taken into account. This reduction, which increases with increasing $T$, is well-known from the virial expansion of the EOS. We can express the reduction introducing the effective cluster energy $E^{\rm eff}_{^2 \rm H}$ as
\begin{equation}
\label{eq:n2H}
n_{^2 \rm H}(T,\mu_n,\mu_p)=\frac{3}{(\pi \hbar^2/mT)^{3/2}}\exp\left(-\frac{E^{\rm eff}_{^2 \rm H }-\mu_n-\mu_p}{T}\right)~, 
\end{equation}
which now becomes temperature and density dependent. Within a more detailed description, the cluster quasiparticle energy is also depending on the center of mass momentum $P$ so that the integral over $P$ is not expressed by the thermal wave length but has to be performed explicitly. In the low-density limit, using the $n - p$ scattering data, the values for the second virial coefficient $b_{pn}(T)$ are given in Ref.~\cite{Horowitz:2006}. The effective binding energy follows as
\begin{equation}
B^{\rm eff}_{^2 \rm H}(T) = -E_{^2 \rm H}^{\rm eff}(T) = -T \ln\left[\frac{\sqrt{2}}{3}b_{pn}(T)\right]
\end{equation}

Similarly, $^4$H and $^5$He can be treated this way. In both cases, no bound states exist, and only the integral over the scattering states remains in Eq. (\ref{gBUd}). The chemical potentials contain the number of neutrons and protons, respectively, of the cluster, and the bound state energies of the scattered ($^4$H, $^4$He) must be added. An effective  cluster energy $E^{\rm eff}_{A,Z}$ can be introduced which is also strongly dependent on $T$, but also on the density. Details are given in Ref.~\cite{Roepke:2020}, where also analytical expressions for the $T,\rho$ dependence of  $E^{\rm eff}_{A,Z}$ for $^4$H and $^5$He are given. These quantities are shown in Fig.1. In the low-density limit, for $^5$He we consider,
\begin{equation}
b_{\alpha n}(T)=\left(\frac{5}{4} \right)^{3/2} \frac{1}{\pi T} \int_0^\infty dE e^{-E/T} \delta_{\alpha n}(E)
\end{equation}
which is also considered in Ref.~\cite{Horowitz:2006}. The effective cluster energy follows as
\begin{equation}
E^{\rm eff}_{^5 \rm He}(T) = E_{^5 \rm He} - T \ln\left[\frac{4}{5^{3/2}}b_{\alpha n}(T)\right]~.
\end{equation}
For $^4$H, the virial expression and the corresponding effective cluster energy in the low-density limit follow from the $^3 {\rm H} - n$ scattering data~\cite{OConnor:2007}.

\section{Weak processes involving light clusters}\label{sec:Appendix_c}
The novel weak reactions involving light clusters with $1<Z<2$ are listed in Table~\ref{tab:nu-reactions-light}. In order to include this set of charged current weak processes into the collision integral of the Boltzmann equation of {\tt AGILE-Boltztran},
\begin{equation}
\left.\frac{1}{c}\frac{d f_\nu}{dt}\right\vert_{\rm collisions}(E_\nu,\mu) = j_\nu(E_\nu)\left(1-f_\nu(E_\nu,\mu)\right) - \chi_\nu(E_\nu) f_\nu(E_\nu,\mu)~,
\end{equation}
with incoming neutrino distribution $f_\nu(E_\nu,\mu)$ depending on the incoming neutrino energy $E_\nu$ as well as on the neutrino scattering angle $\mu=\sin\vartheta$ (for illustration, see Fig.~1 in Ref.~\cite{Mezzacappa:1993gm}), the total neutrino emissivity $j_\nu$ and opacity $\chi_\nu$ are supplemented by those of the reactions listed in Table~\ref{tab:nu-reactions-light},
\begin{subequations}
\begin{equation}
j_{\nu_e} = j_{\nu_e n} + j_{\nu_e \, ^2 {\rm H}} + j_{\nu_e \, nn} + j_{\nu_e \, ^3 {\rm H}} + j_{\nu_e \, ^4 {\rm H}}~,\;\;\;\;
\chi_{\nu_e} = \chi_{\nu_e n} + \chi_{\nu_e\, ^2 {\rm H}} + \chi_{\nu_e \, nn} + \chi_{\nu_e \, ^3 {\rm H}} + \chi_{\nu_e \, ^4 {\rm H}}~,
\end{equation}
\begin{equation}
j_{\bar\nu_e} = j_{\bar\nu_e p} + j_{\bar\nu_e \, ^2 {\rm H}} + j_{\bar\nu_e \, pp} + j_{\bar\nu_e \, ^3 {\rm He}} + j_{\bar\nu_e \, ^4 {\rm He}}~,\;\;\;\;
\chi_{\bar\nu_e} = \chi_{\bar\nu_e n} + \chi_{\bar\nu_e ^2 {\rm H}} + \chi_{\bar\nu_e \, pp} + \chi_{\bar\nu_e \, ^3 {\rm He}} + \chi_{\bar\nu_e \, ^4 {\rm He}}~,
\end{equation}
\end{subequations}
where the individual emissivities and opacities are related via the relation of detailed balance (Appendix~\ref{sec:Appendix_c5}). In the following sections the derivation of these reaction rates is provided.

\subsection{Absorption of $\nu_e$ and $\bar\nu_e$ on $^2$H}\label{sec:Appendix_c1}
For the following weak processes with deuteron,
\begin{equation}
\nu_e+ {}^2\text{H} \rightarrow p+p+e^- \;,\qquad
\bar\nu_e+ {}^2\text{H} \rightarrow n + n + e^+ \;, 
\end{equation}
we can introduce relative and center-of-mass momenta,
\begin{equation}
{\bf P} = {\bf p}_1 + {\bf p}_2\;,\qquad {\bf p} = \frac{{\bf p}_1-{\bf p}_2}{2}\;,
\end{equation}
with individual nucleon momenta, ${\bf p}_1$ and ${\bf p}_2$, and express energy and momentum conservation in these coordinates as follows,
\begin{equation}
{\bf p}_\nu + {\bf p}_{^2\text{H}} = {\bf p}_e + {\bf P}~,
\qquad
E_\nu + \frac{{\bf p}^2_\nu}{2m_{^2\text{H}}} + m_{^2\text{H}} = E_e + \frac{{\bf P}^2}{4m_p} + \frac{{\bf p}^2}{m_p} + 2m_p~.
\end{equation}
Here we replace the effective masses with the vacuum masses since $1/m_{^2\text{H}}^*\simeq1/m_{^2\text{H}}$ and $1/m_N^*\simeq 1/m_N$. In the calculations of the differential cross sections in Ref.~\cite{Nakamura:2001}, the authors neglects the dependence of the cross section on the direction of ${\bf p}$, they compute an angle-averaged value. For the neutrino transport, we need to recover the dependence of the cross-section on the direction of ${\bf p}$. For convenience we will define it with respect to the direction of the vector ${\bf P}$, for which we define the angle $\theta$ such that ${\bf p}\cdot{\bf P}=pP\cos\theta$ and the differential cross section is given as follows,
\begin{equation}
\label{eq:3dto2d}
\frac{d\sigma_{\nu\,^2\text{H}}}{d(\cos\theta)d\Omega_edp_e}(E_\nu)= \frac{1}{2}\frac{d\sigma_{\nu\,^2\text{H}}}{d\Omega_edp_e}(E_\nu)~,
\end{equation}
with the assumption of isotropy. 

The general expression for the reaction rates for the charged current absorption processes with deuteron, break-up reactions (1) and (2) in Table~\ref{tab:nu-reactions-light}, is given by the following integral equation,
\begin{eqnarray}
\label{eq:cc_deuteron_full}
\chi_{\nu\,^2{\rm H}}({E_\nu}) &=&
\frac{2\pi G^2}{\hbar\,c} g_{^2\text{H}} \int
\frac{d^3p_{^2\text{H}}}{(2\pi\hbar c)^3}
\frac{d^3p_e}{(2\pi\hbar c)^3}
\frac{d^3p_1}{(2\pi\hbar c)^3}
\frac{d^3p_2}{(2\pi\hbar c)^3}
\,
\tilde{f}_{^2\text{H}}(E_{^2\text{H}})
\left(1-f_e(E_e)\right)
\left(1-f_1(E_1)\right)
\left(1-f_2(E_2)\right)
\\
&&
\qquad\times~
(2\pi\hbar c)^3\,\left\vert M \right\vert^2\,\delta^4\left(p_\nu+p_{^2\text{H}}-p_e-p_1-p_2\right)\;,
\nonumber
\end{eqnarray}
with the product of the Fermi coupling constant, $G_F$, and the $ud$ entry of
the Cabibbo-Kobayashi-Maskawa matrix, denoted as $G=G_F V_{ud}$, with the Fermi-Dirac
equilibrium distribution functions $f_i$, except the Bose-Einstein
distribution function $\tilde f_{^2 \rm H}$ for deuteron, as well as
with spin averaged interaction matrix element $M$ which connects
initial and final states. In the zero-momentum exchange approximation,
${\bf q}={\bf p_\nu} - {\bf p}_e$ with $q\simeq 0$,
expression~\eqref{eq:cc_deuteron_full} reduce to the following form,
\begin{eqnarray}
\label{eq:nuedopa}
  \chi_{\nu\,^2{\rm H}}(E_\nu) &=&
\frac{2\pi G^2}{\hbar\,c} g_{^2\text{H}} \int
\frac{d^3p_{^2\text{H}}}{(2\pi\hbar c)^3}
\frac{d^3p_e}{(2\pi\hbar c)^3}
\frac{d^3P}{(2\pi\hbar c)^3}
d^3p
\,\left\vert M \right\vert^2
\tilde{f}_{^2\text{H}}(E_{^2\text{H}})
\left(1-f_e(E_e)\right)
\left(1-f_1(E_1)\right)
\left(1-f_2(E_2)\right)
\\
&&
\qquad
\times~
\delta\left(E_\nu+E_{^2\text{H}}-E_e-E_1-E_2\right)
\delta^3\left({\bf p}_{^2\text{H}} - {\bf P}\right)
\nonumber
\\
\\
&=&
\frac{2\pi G^2}{\hbar\,c(2\pi\hbar c)^3} g_{^2\text{H}} \int
\frac{d^3p_{^2\text{H}}}{(2\pi\hbar c)^3}
\frac{d^3p_e}{(2\pi\hbar c)^3}
2\pi p^2\,d(\cos\theta)
\,\left\vert M \right\vert^2
\tilde{f}_{^2\text{H}}(E_{^2\text{H}})
\left(1-f_e(E_e)\right)
\left(1-f_1(E_1)\right)
\left(1-f_2(E_2)\right)\;,
\label{eq:cc_deuteron_reduced}
\end{eqnarray}
with
\begin{subequations}
\begin{equation}
{\bf P} = {\bf p}_{^2\text{H}}\;,\qquad
p = \sqrt{m_p\left(E_\nu-E_e + m_{^2\text{H}} - 2m_p\right)}\;,
\label{eq:momentaa}
\end{equation}
\begin{equation}
E_1 = \frac{1}{2m_p}\left(\frac{p_{^2\text{H}}^2}{4}+p^2+pp_{^2\text{H}}\cos\theta\right) + m_p\;,\qquad
E_2 = \frac{1}{2m_p}\left(\frac{p_{^2\text{H}}^2}{4}+p^2-pp_{^2\text{H}}\cos\theta\right) + m_p\,.
\label{eq:momentac}
\end{equation}
\end{subequations}
In order to relate the opacity with the double-differential cross
section computed in Ref.~\cite{Nakamura:2001} we use the following
association:
\begin{equation}
  \frac{d\sigma_{\nu\,^2\text{H}}}{d(\cos\theta)d\Omega_edp_e}(E_\nu) = \frac{2\pi\, G^2}{(\hbar c)(2\pi\hbar c)^6}2\pi \left(p_e p\right)^2 \left\vert M \right\vert^2\;,
\end{equation}
and obtain the following expression for the opacity,
\begin{equation}
\chi_{\nu\,^2{\rm H}}(E_\nu) =
g_{^2\text{H}}
\int
\frac{d^3p_{^2\text{H}}}{(2\pi \hbar c)^3}
d\Omega_e dp_e d(\cos\theta)
\left(\frac{d\sigma_{\nu\,^2\text{H}}}{d(\cos\theta)d\Omega_e dp_e}(E_\nu^*)\right)
\tilde{f}_{^2\text{H}}(E_{^2\text{H}})
(1-f_e(E_e))
\left(1-f_1(E_1)\right)
\left(1-f_2(E_2)\right)\;,
\end{equation}
with the medium-modified cross section due to the shifted neutrino
energy $E_\nu^*$, which can be obtained via the following equations,
\begin{equation}
E_{\nu_e}^*=E_{\nu_e}+(m^*_{^2\text{H}}-m_{^2\text{H}})+U_{^2\text{H}}-2(m_p^*-m_p)-2U_p~,
\label{eq:Enue_shift_deuteron}
\end{equation}
\begin{equation}
E_{\bar\nu_e}^*=E_{\bar\nu_e}+(m^*_{^2\text{H}}-m_{^2\text{H}})+U_{^2\text{H}}-2(m_n^*-m_n)-2U_n~.
\label{eq:Enueb_shift_deuteron}
\end{equation}
The medium effects, i.e. the mean-field potentials and the effective masses, are provided by the gRDF(DD2) EOS. They also modify Eqs.~\eqref{eq:momentaa}--\eqref{eq:momentac} as follows,
\begin{subequations}
\begin{equation}
p = \sqrt{m_p^*\left(E_\nu-E_e + m_{^2\text{H}}^* - 2m_p^*+U_d-2U_p\right)}\;,
\end{equation}
\begin{equation}
E_1 = \frac{1}{2m_p^*}\left(\frac{p_{^2\text{H}}^2}{4}+p^2+pp_{^2\text{H}}\cos\theta\right) + m_p^* + U_p\;,\qquad
E_2 = \frac{1}{2m_p^*}\left(\frac{p_{^2\text{H}}^2}{4}+p^2-pp_{^2\text{H}}\cos\theta\right) + m_p^* + U_p\,.
\end{equation}
\end{subequations}
Furthermore, since final state Pauli-blocking does not depend on the
electron angular direction one can integrate the electron solid angle
and obtain the final expression for the opacity for (anti)neutrino
absorption on $^2$H,
\begin{equation}
\chi_{\nu\,^2{\rm H}}(E_\nu) =
\frac{g_{^2\text{H}}}{2}
\int
\frac{d^3p_{^2\text{H}}}{(2\pi \hbar c)^3}
dp_e \, d(\cos\theta)
\left(\frac{d\sigma_{\nu\,^2\text{H}}}{dp_e}(E_\nu^*)\right)
\tilde{f}_{^2\text{H}}(E_{^2\text{H}})
\,(1-f_e(E_e))
\left(1-f_1(E_1)\right)
\left(1-f_2(E_2)\right)\;,
\label{eq:cc_ve_deuteron}
\end{equation}
where we have used Eq.~\eqref{eq:3dto2d}. The remaining three integrals are evaluated computationally on flight in the SN simulation reported above in Sec.~\ref{sec:v-reactions}.
%We therefore further reduce Eq.~\eqref{eq:cc_deuteron_sigma} neglecting final state nucleon blocking. The deuteron integration factorizes and one is left with only one integration over the electron momentum,
%
%\begin{eqnarray}
%&&1/\lambda(E_\nu) =
%n_{^2\text{H}}
%\int
%dp_e 
%\left(\frac{d\sigma_{\nu\,^2\text{H}}}{dp_e}(E_\nu^*)\right)
%(1-f_e(E_e))~.
%\label{eq:cc_deuteron_proxy}
%\end{eqnarray}
%
%This approximation is valid in the limit of low nucleon degeneracy, which is always applies for protons in the final state ($\nu_e$-channel). It is also fulfilled for neutrons in the final state at low and intermediate densities ($\bar\nu_e$-channel). However, not that  expression~\eqref{eq:cc_deuteron_proxy} does not fulfill detailed balance.
%

\subsection{$e^\pm$ captures on $^2$H}\label{sec:Appendix_c2}

In addition to the (anti)neutrino absorption on deuteron, we can also consider electron captures on deuteron,
\begin{equation}
  e^- +{} ^2\text{H} \rightarrow n + n + \nu_e \;,\qquad
  e^+ + {}^2\text{H} \rightarrow p + p + \bar\nu_e 
\end{equation}
for which the differential cross section is given by the following expression,
\begin{equation}
\frac{d\sigma_{e\,^2\text{H}}}{d\Omega_{\nu_e} dp_{\nu}}(E_e) = \frac{2\pi G^2}{2(\hbar c)(2\pi\hbar c)^6}4\pi p_{\nu}^2 p(E_e,E_{\nu})^2 \left\vert M \right\vert^2\;,
\label{eq:sigma_deuteron_electron}
\end{equation}
with additional factor 2 due to the two internal degrees of freedom
for the electron. In the following we will examine the reaction with
electrons, for which we have the following settings,
\begin{subequations}
\begin{equation}
{\bf P} = {\bf p}_{^2\text{H}}\;,\qquad
p(E_e,E_{\nu_e}) = \sqrt{m_n\left(E_e-E_{\nu_e} + m_{^2\text{H}} - 2m_n\right)}\;,
\end{equation}
\begin{equation}
E_1 = \frac{1}{2m_n}\left(\frac{p_{^2\text{H}}^2}{4}+p^2+pp_{^2\text{H}}\cos\theta\right) + m_n\;,\qquad
E_2 = \frac{1}{2m_n}\left(\frac{p_{^2\text{H}}^2}{4}+p^2-pp_{^2\text{H}}\cos\theta\right) + m_n\,.
\end{equation}
\end{subequations}
Furthermore, note that cross section~\eqref{eq:sigma_deuteron_electron} can be related with the cross section for $\bar\nu_e$-absorption on deuteron (reaction~(2) in Table~\ref{tab:nu-reactions-light}) as follows,
\begin{equation}
\frac{d\sigma_{\bar\nu_e\,^2\text{H}}}{d\Omega_{e} dp_e}(E_{\bar\nu_e}) = \frac{2\pi G^2}{(\hbar c)(2\pi\hbar c)^6}4\pi p_e^2 p(E_{\bar\nu_e},E_e)^2 \left\vert M \right\vert^2\;,
\end{equation}
with
\begin{subequations}
\begin{equation}
{\bf P} = {\bf p}_{^2\text{H}}\;,\qquad
p(E_{\bar\nu_e},E_e) = \sqrt{m_n\left(E_{\bar\nu_e}-E_e + m_{^2\text{H}} - 2m_n\right)}\;,
\end{equation}
\begin{equation}
E_1 = \frac{1}{2m_n}\left(\frac{p_{^2\text{H}}^2}{4}+p^2+pp_{^2\text{H}}\cos\theta\right) + m_n\;,\qquad
E_2 = \frac{1}{2m_n}\left(\frac{p_{^2\text{H}}^2}{4}+p^2-pp_{^2\text{H}}\cos\theta\right) + m_n\,.
\end{equation}
\end{subequations}
Here we find the following relationship between the cross section for capturing an electron of energy $E_e$ producing a $\nu_e$ of momentum $p_{\nu_e}$ and the cross section for $\bar\nu_e$-absorption of the same energy, $E_{\bar\nu_e} = E_e$ and producing a positron of momentum $p_e=p_{\bar\nu_e}$:
\begin{equation}
  \label{eq:ectonuebar}
\frac{d\sigma_{e^-\,^2\text{H}}}{d\Omega_{\nu_e} dp_{\nu_e}}(E_e) =
\frac{1}{2}\frac{p(E_e,p_{\bar\nu_e})^2}{p(E_{e},E_e(p_{\bar\nu_e}))^2}
\frac{d\sigma_{\bar\nu_e\,^2\text{H}}}{d\Omega_e dp_e}(E_e) 
\simeq 
\frac{1}{2}
\frac{d\sigma_{\bar\nu_e\,^2\text{H}}}{d\Omega_e dp_e}(E_e)\;,
\end{equation}
assuming relativistic electrons. The $\nu_e$-emissivity can be
obtained from equation~\eqref{eq:nuedopa} using the change
$(1-f_e)\rightarrow 2 f_e$  to obtain after substituting the
electron-capture cross section:
\begin{equation}
j_{e^-{}^2\text{H}}(E_{\nu_e}) =
\frac{g_{^2\text{H}}}{2}
\int
\frac{d^3p_{^2\text{H}}}{(2\pi\hbar c)^3}
d(\cos\theta) d\Omega_e dp_e 
\left(\frac{p_e^2}{p_{\nu_e}^2}\frac{d\sigma_{e^-\,^2\text{H}}}{d\Omega_{\nu_e}
    dp_{\nu_e}}(E_e^*)\right) 
\,\tilde{f}_{^2\text{H}}(E_{^2\text{H}})
\,2\,f_e(E_e)\,(1-f_1(E_1))\,(1-f_2(E_2))\;.
\label{eq:ve-d-emissivity}
\end{equation}
Applying now detailed balance, $1/\lambda(E_{\nu_e}) =
e^{\beta(E_{\nu_e}-\mu_{\nu_e}^\text{eq})}\,j(E_{\nu_e})$ (for
details, see Appendix~\ref{sec:Appendix_c5}), and expressing the
electron capture cross section by means of the neutrino-cross section, see
eq.~\eqref{eq:ectonuebar}, 
computed in Ref.~\cite{Nakamura:2001} one obtains the opacity,
\begin{equation}
\chi_{\nu_e\,nn}({E_{\nu_e}}) =
\frac{g_{^2\text{H}}}{2}
\int
\frac{d^3p_{^2\text{H}}}{(2\pi \hbar c)^3}
d(\cos\theta)
d\Omega_e
dp_e
\left(\frac{p_e^2}{p_{\nu_e}^2}
\frac{d\sigma_{\bar\nu_e\,^2\text{H}}}{d\Omega_e dp_{\nu_e}}(E_e^*)\right)
\left(1+\tilde{f}_{^2\text{H}}(E_{^2\text{H}})\right)
\left(1-f_e(E_e)\right)
\,f_1(E_1)\,2\,f_2(E_2)~,
\end{equation}
with medium-modified electron energy as follows,
\begin{equation}
E_e^* = E_e + (m_{^2\text{H}}^*-m_{^2\text{H}}) - 2 (m_n^*-m_n) + (U_d-2U_n)\;,
\label{eq:Enu_shift_deuteron_electron}
\end{equation}
and
\begin{subequations}
\begin{equation}
p = \sqrt{m_n\left(E_e-E_{\nu_e} + m_{^2\text{H}}^* - 2m_n^* + U_d-2U_n\right)}\;,
\end{equation}
\begin{equation}
E_1 = \frac{1}{2m_n^*}\left(\frac{p_{^2\text{H}}^2}{4}+p^2+pp_{^2\text{H}}\cos\theta\right) + m_n^* + U_n\;,\qquad
E_2 = \frac{1}{2m_n^*}\left(\frac{p_{^2\text{H}}^2}{4}+p^2-pp_{^2\text{H}}\cos\theta\right) + m_n^* + U_n\,.
\end{equation}
\end{subequations}
Furthermore, we can integrate over the electron solid angle and recover the following expression,
\begin{equation}
\chi_{\nu\,N_1 N_2}({E_{\nu_e}}) =
\frac{g_{^2\text{H}}}{4}
\int
\frac{d^3p_{^2\text{H}}}{(2\pi \hbar c)^3}
dp_e d(\cos\theta)
\left(\frac{p_e^2}{p_{\nu_e}^2}
\frac{d\sigma_{\bar\nu_e\,^2\text{H}}}{dp_{\nu_e}}(E_e^*)\right)
\left(1+\tilde{f}_{^2\text{H}}(E_{^2\text{H}})\right)
\left(1-f_e(E_e)\right)
\,2\,f_1(E_1)\,2\,f_2(E_2)~.
\label{eq:cc_ec_deuteron}
\end{equation}
%
%In the spirit of \eqref{eq:cc_deuteron_proxy}, in the presence of low degeneracy Eq.~\eqref{eq:ve-d-emissivity} reduces as follows,
%
%\begin{eqnarray}
%j({E_{\nu_e}}) =
%\frac{n_{^2\text{H}}}{2}
%\int d p_e
%\left(
%\frac{p_e^2}{p_\nu^2}
%\frac{d\sigma_{\bar\nu_e\,^2\text{H}}}{dp_{\nu_e}}(E_e^*)
%\right) f_e(E_e)\;.
%\label{eq:cc_deuteron_electron_proxy}
%\end{eqnarray}
%
%An equivalent expression is obtained for the $\bar\nu_e$-emissivity from positron captures on deuteron. Note here also that expression~\eqref{eq:cc_deuteron_electron_proxy} does not fulfill detailed balance.
%

\end{widetext}

\subsection{$\nu_e$ absorption on $^3$H and $\bar\nu_e$ absorption on $^3$He}\label{sec:Appendix_c3}

For charged current $\nu_e$-absorption on $^3$H the situation is somewhat more complicated than for $^2$H, because there are two components. The first one is the transition to the bound state of $^3$He and the second component is the transition into the continuum with three final-state nucleons:
\begin{equation}
\nu_e + {}^3\text{H} \rightarrow \,
^3\text{He} + e^-\;,\quad
\nu_e + {}^3\text{H} \rightarrow \,
p + p + n + e^-\;, 
\label{eq:3H-3He}
\end{equation}
For the bound state to bound state reaction, we obtain the opacity:

\begin{equation}
\chi_{\nu_e \, ^3 {\rm H}}({E_{\nu_e}}) \simeq n_{^3\text{H}} \,\sigma_{\nu_e \,^3 \rm H}(E_{e^-})\,\left(1-f_{e^-}(p_{e^-})\right)~,
%n_{^3\text{H}} \frac{G^2\,V_{ud}^2}{\pi(\hbar\,c)^4}\,p_{e^-}\,E_{e^-}\,\left(1-f_{e^-}(E_{e^-})\right) \rm B(GT)~,
\label{eq:cc_ve_triton}
\end{equation}
neglecting final-state $^3$He blocking where we introduced
the triton number density $n_{^3\text{H}}$ and the 
corresponding cross section,
\begin{eqnarray}
\sigma_{\nu_e \,^3 \rm H}(E_{e^-}) &=& \frac{G^2}{\pi}\frac{1}{(\hbar\,c)^4}
\left(p_{e^-}E_{e^-}\right) \, \mathcal{B_{\rm GT}} 
\label{eq:sigma_triton}
\\
&\propto&
2.26~\times 10^{-41}\,\left(\frac{E_{\nu_e}}{15~{\rm MeV}}\right)\,\rm cm^2~ \rm MeV^{-2}~,
\nonumber
\end{eqnarray}
for a typical neutrino energy of 15~MeV, and with the Gammow-Teller transition amplitude, $\mathcal{B}_{\rm GT}=5.87$, which is known from the triton decay. The following relation connects the medium modified electron energy relation,
\begin{equation}
E_{e^-} = E_{\nu_e} + \left(m_{^3 \rm H} - m_{^3 \rm He}\right) + \left(U_{^3 \rm H} - U_{^3 \rm He}\right)~,
\end{equation}
with the neutrino energy $E_{\nu_e}$ including the mass difference between $^3$H and  $^3$He, $m_{^3 \rm H} - m_{^3 \rm He}=0.5296$ (including electron contributions), and the mean-field potentials, $U_{^3 \rm H}$ and $U_{^3 \rm He}$. The latter quantities are provided by the gRDF(DD2) EOS. 

\begin{figure}
\centering
\includegraphics[width=1\columnwidth]{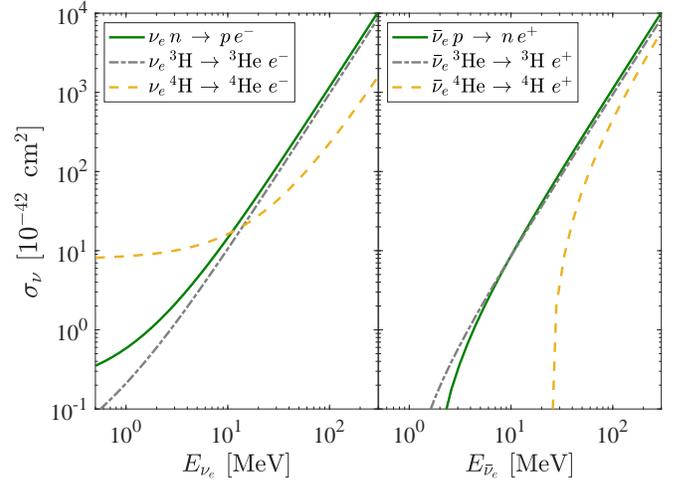}
\caption{Neutrino absorption cross section for reactions involving $^3$H and $^3$He (grey dash-dotted lines) as well as $^4$H and $^4$He (yellow dashed lines), in comparison to those for the charged current reactions~(1)--(3) in Table~\ref{tab:nu-reactions}, for $\nu_e$ (left panel) and $\bar\nu_e$ (right panel).}
\label{fig:sigma_light}
\end{figure}

From the comparison of the cross section of the charged current reactions~(1)--(3) in Table~\ref{tab:nu-reactions}, for which $\sigma_{\nu_e N}\simeq 2.47 \times 10^{-41}$~cm$^2$~MeV$^{-2}$, again for a typical neutrino energy of 15~MeV, and with the transition standard amplitude, $g_V^2 + 3 g_A^2$, with vector and axial-vector coupling constants, $g_V=1.0$ and $g_A=1.27$, it becomes evident that reactions involving $^3$H and $^3$He can become as important as the reactions involving the unbound nucleons, depending on the kinematics including final-state blocking contributions. This has already been realised in Ref.~\cite{Fischer:2017} at the level of the cross section analysis. Figure~\ref{fig:sigma_light} compares these cross section for $\nu_e$ (left panel) and $\bar\nu_e$ (right panel). We omit here the $^3$H cross sections for the break-up reactions and the reactions involving $^2$H (see therefore Ref.~\cite{Fischer:2017}). 

The second component of the $\nu_e$-absorption on triton in
\eqref{eq:3H-3He} is to transitions to the continuum, which is
negligible compared to the bound state to bound state reaction as was
shown in Ref.~\cite{Fischer:2017} based on the cross sections calculated
in analogy to Ref.~\cite{Arcones:2008}.

A similar expression as \eqref{eq:cc_ve_triton} is obtained for the opacity of $\bar\nu_e$-absorption on $^3$He,
\begin{equation}
\bar\nu_e + {}^3\text{He} \rightarrow \, ^3\text{H} + e^+\;\;,
\end{equation}
as follows, 
\begin{equation}
\chi_{\bar\nu_e \, ^3 {\rm He}}({E_{\bar\nu_e}}) \simeq n_{^3\text{He}} \,\sigma_{\bar\nu_e \,^3 \rm He}(E_{\bar\nu_e})\,\left(1-f_{e^+}(p_{e^+})\right)
%n_{^3\text{He}} \frac{G^2\,V_{ud}^2}{\pi(\hbar\,c)^4}\,p_{e^+}\,E_{e^+}\,\left(1-f_{e^+}(E_{e^+})\right) {\rm B(GT)}\;,
\label{eq:cc_veb_triton}
\end{equation}
The cross section, $\sigma_{\bar\nu_e\,^3 \rm He}$, involves the same
matrix element as \eqref{eq:sigma_triton}, and the following medium
modified positron energy relation,
\begin{equation}
E_{e^+} = E_{\bar{\nu}_e} - \left(m_{^3 \rm H} - m_{^3 \rm He}\right) - \left(U_{^3 \rm H} - U_{^3 \rm He}\right)~,
\end{equation}
again with $^3$H and $^3$He masses and their mean-field potentials. Note that expression~\eqref{eq:cc_veb_triton} also neglects final state triton contributions. Also here the transition to the continuum, $\bar\nu_e\,^3\text{H} \longrightarrow n\,n\,n\,e^+$, is negligible as was shown in Ref.~\cite{Fischer:2017}, based on the cross section similar to those provided in Ref.~\cite{Arcones:2008}.

\subsection{$\bar\nu_e$ absorption on $^4$He and $\nu_e$ absorption on $^4$H}\label{sec:Appendix_c4}
For the anti-neutrino charged current channel, there is the following bound state reaction:
\begin{equation}
\bar\nu_e + {}^4\text{He} \rightarrow {} ^4\text{H} + e^+~.
%\bar\nu_e\,^4\text{He} \longrightarrow \, ^3 {\rm H} \,n \,  e^+~.
\label{eq:reaction_4He}
\end{equation}

The description of the final state $^4$H is very challenge due 
to the presence of several broad resonances and the fact that 
the presence of a degeneate neutron background suppresses the 
emission of neutrons in the medium. Here, in a first attempt 
to take this process into account, we use the 
thermally-averaged  inclusive cross sections, $\langle \sigma 
\rangle_T$, that have been provided in Ref.~\cite{Gazit:2007PhRvL}, 
assuming a thermal neutrino distribution
with zero chemical potential, and assume that in the medium 
all neutron emission channels are suppressed. In order to be 
able to implement them into the collision integral of 
{\tt AGILE-BOLTZTRAN}, we reverse engineer in order to obtain 
the transition amplitude as follows,
\begin{equation}
\langle \sigma \rangle_T = \frac{4\pi}{(hc)^3}\frac{1}{n_\nu}\int dE_{\bar\nu_e} E_{\bar\nu_e}^2\, \sigma_{\bar\nu_e\,^4{\rm He}}(E_{e^+}) \,f_{\bar\nu_e}^{\mu_{\bar\nu_e}=0}(E_{\bar\nu_e};T)~,
\end{equation}
with neutrino number density, $n_\nu$, and we set
\begin{equation}
\sigma_{\bar\nu_e\,^4{\rm He}}(E_{e^+}) = \frac{G^2}{\pi}\frac{1}{(\hbar c)}\left(p_{e^+} E_{e^+}\right)\,\mathcal{B}_{\rm GT}~,
\end{equation}
assuming a constant value of $\mathcal{B}_{\rm GT}=4.75$. This
approximation is fitted to reproduce the values in 
Ref.~\cite{Gazit:2007PhRvL} in the temperature range of 
$T\simeq 5-10$~MeV (see Table~\ref{tab:sigma}) within a factor 
of 2 or less. However, it certainly overestimates the 
thermally-averaged cross sections at low temperatures by about 
an order of magnitude. Similarly as before, the anti-neutrino 
and positron energies are related via the restmass and 
mean-field potential differences of $^4$H and $^4$He as 
follows,
\begin{equation}
E_{e^+} = E_{\bar\nu_e} - \left(m_{^4 \rm H} - m_{^4 \rm He}\right) - \left(U_{^4 \rm H} - U_{^4 \rm He}\right)~.
\end{equation}
Then, the opacity for \eqref{eq:reaction_4He} in the elastic 
approximation reads as follows,
\begin{equation}
\chi_{\bar\nu_e\,^4 {\rm He}}({E_{\bar\nu_e}}) = n_{^4 \rm He}\,\sigma_{\bar\nu_e\,^4{\rm He}}(E_{e^+}) \left(1-f_{e^+}(E_{e^+})\right)~,
\label{eq:cc_ve_4he}
\end{equation}
with the $^4$He number density denoted as $n_{^4{\rm He}}$.

\begin{table}
\centering
\caption{Thermally average cross section for \eqref{eq:reaction_4He}, comparing our fit with the data of Ref.~\cite{Gazit:2007PhRvL}.}
\begin{tabular}{ccc}
\hline
\hline
$T$ & $\langle \sigma \rangle_T$ & Ref. \\
$[$MeV$]$ & $[10^{-42}$~cm$]$ & $[10^{-42}$~cm$]$ \\
\hline
2 & $9.33(-4)$ & $1.42(-4)$ \\
4 & $3.78(-1)$ & $1.80(-1)$ \\
5 & $1.38$ & --- \\
6 & $3.41$ & $3.70$ \\
8 & $11.61$ & $22.9$ \\
\hline
\end{tabular}
\label{tab:sigma}
\end{table}

The matrix element entering in the cross section above for neutrino absorption on $^4$H,
\begin{equation}
\nu_e+{}^4\text{H} \longrightarrow\,^4\text{He}+ e^-
\label{eq:reaction_4H}
\end{equation}
is generally not known for this reaction in vacuum and furthermore should be modified due to the in-medium effects. However, assuming mirror symmetry between $^4$H and $^4$Li, it is taken to be a factor of 5 smaller than the one for \eqref{eq:reaction_4He}, which yields $\mathcal{B}_{\rm GT}=0.95$. Note that here we consider only transitions to bound states in $^{4}$He as they are energetically favored. Similarly as for the processes involving $^3$H and $^3$He in Sec.~\ref{sec:Appendix_c3}, we express the reaction rate for \eqref{eq:reaction_4H} as follows,
\begin{equation}
\chi_{\nu_e\,^4 {\rm H}}({E_{\nu_e}}) = n_{^4 \rm H}\,\sigma_{\nu_e \,^4 \rm H}(E_{e^-})\, \left(1-f_{e^-}(E_{e^-})\right)~,
\label{eq:cc_ve_4h}
\end{equation}
with $^4$H number density, $n_{^4{\rm H}}$, and the following medium modified relation between electron and neutrino energies,
\begin{equation}
E_{e^-} = E_{\nu_e} + \left(m_{^4 \rm H} - m_{^4 \rm He}\right) + \left(U_{^4 \rm H} - U_{^4 \rm He}\right)~.
\end{equation}
The cross section,
\begin{equation}
\sigma_{\nu_e\,^4{\rm H}}(E_{e^-}) = \frac{G^2}{\pi}\frac{1}{(\hbar c)}\left(p_{e^-} E_{e^-}\right)\,\mathcal{B}_{\rm GT}~,
\end{equation}
is also shown in Fig.~\ref{fig:sigma_light} in the left panel (yellow dashed line).

\subsection{Detailed balance}\label{sec:Appendix_c5}
Here we will derive the detailed balance relation for the (anti)neutrino emissivity $j(E_\nu)$, i.e. the reverse reactions,
\begin{equation*}
e^- + p + p \longrightarrow \,^2\text{H} +\nu_e \;,\qquad e^+ + n + n \longrightarrow \,^2\text{H} + \bar\nu_e \;,
\end{equation*}
for which the reaction rate is given by the following expression,
\begin{widetext}
\begin{eqnarray}
j_{\nu_e\,^2{\rm H}}(E_\nu)
=
\frac{2\pi G^2}{\hbar c}\frac{1}{(2\pi\hbar c)^3} 
\int
\frac{d^3p_{^2\text{H}}}{(2\pi\hbar c)^3}
\frac{d^3p_e}{(2\pi\hbar c)^3}
2\pi p^2\,d(\cos\theta)
\,\left\vert M \right\vert^2
\left\{1 + \tilde{f}_{^2\text{H}}(E_{^2\text{H}})\right\}
f_e(E_e)\,f_1(E_1)\,f_2(E_2)~.
\end{eqnarray}
Comparing this expression with \ref{eq:cc_deuteron_reduced} one obtains,
\begin{eqnarray}
&&
j_{\nu_e\,^2{\rm H}}(E_\nu) =
\frac{2\pi G^2}{\hbar c}\frac{1}{(2\pi\hbar c)^3}
\int
\frac{d^3p_{^2\text{H}}}{(2\pi\hbar c)^3}
\frac{d^3p_e}{(2\pi\hbar c)^3}
2\pi p^2\,d(\cos\theta)
\,\left\vert M \right\vert^2
\exp\left\{\beta(E_{^2\text{H}}-\mu_{^2\text{H}})\right\}
\tilde{f}_{^2\text{H}}(E_{^2\text{H}})
\,\exp\left\{-\beta(E_e-\mu_e)\right\}\left(1-f_e(E_e)\right)
\nonumber
\\ \nonumber \\
&&
\qquad\qquad\qquad\times\,\,
\exp\left\{-\beta(E_1-\mu_1)\right\}\left(1-f_1(E_1)\right)
\exp\left\{-\beta(E_2-\mu_2)\right\}\left(1-f_2(E_2)\right)
\\ \nonumber \\
&&
\quad
= \frac{2\pi G^2}{\hbar c}\frac{1}{(2\pi\hbar c)^3}
\int
\frac{d^3p_{^2\text{H}}}{(2\pi\hbar c)^3}
\frac{d^3p_e}{(2\pi\hbar c)^3}
2\pi p^2\,d(\cos\theta)
\,\left\vert M \right\vert^2
\,\exp\left\{-\beta(E_{\nu_e}-\mu_{\nu_e}^\text{eq})\right\}
\tilde{f}_{^2\text{H}}(E_{^2\text{H}})
\left(1-f_e(E_e)\right)\left(1-f_1(E_1)\right)\left(1-f_2(E_2)\right)
\nonumber \\ \nonumber \\
&&
\quad
=\exp\left\{-\beta(E_{\nu_e}-\mu_{\nu_e}^\text{eq})\right\}\,\chi_{\nu_e\,^2{\rm H}}(E_{\nu_e})~,
\label{eq:detailed_balance}
\end{eqnarray}
\end{widetext}
with the neutrino equilibrium chemical potential, $\mu_{\nu_e}^\text{eq}=\mu_e+\mu_p+\mu_p-\mu_{^2\text{H}}=\mu_e-(\mu_n-\mu_p)$. Note that this is the same detailed balance relation as for the charged current absorption processes with free nucleons (reactions (1) and (2) in Table~\ref{tab:nu-reactions}). Applying the same approach for the reverse triton opacity, i.e. the emissivities for the  following processes,
\begin{eqnarray*}
&& e^- \,\,^2 {\rm H} \longrightarrow \,p\,p\,\nu_e \;,\quad
e^+ \,\,^2 {\rm H} \longrightarrow \,n\,n\,\bar\nu_e \;,\\
&& e^- \,\,^3\text{He} \longrightarrow \,^3\text{H}\,\nu_e \;,\quad
e^+ \,\, ^3\text{H} \longrightarrow \,^3\text{He}\,\bar\nu_e \;,\\
&& e^- \,\, ^4\text{He} \longrightarrow \,^4\text{H}\,\nu_e \;,\quad
e^+ \,\, ^4\text{H} \longrightarrow \,^4\text{He}\,\bar\nu_e \;,
\end{eqnarray*}
one will recover the same detailed balance relation~\eqref{eq:detailed_balance}.

%\bibliography{references}
%

\end{document}